\documentclass[twocolumn]{IEEEtran}

\usepackage{cite}

\usepackage{algorithm}
\usepackage{algpseudocode}
\usepackage{tabularx} 

\usepackage{float}

\usepackage{graphicx}
\usepackage[cmex10]{amsmath}
\usepackage{epstopdf}
\usepackage{cite}
\usepackage{color,soul}
\usepackage{algorithm}
\usepackage{algpseudocode}
\usepackage{caption}
\usepackage{algpseudocode}
\usepackage{bm}
\usepackage{graphicx}
\usepackage{subcaption}
\DeclareMathOperator\erf{erf}
\usepackage{amssymb}

\usepackage{amsmath}
\usepackage{stfloats}
\usepackage{blindtext}

\usepackage{graphicx}
\usepackage{subcaption}

\usepackage[table]{xcolor}

\definecolor{myGreen1}{rgb}{0.9098, 0.9529, 0.8824}

\begin{document}

\setlength{\abovedisplayskip}{2pt}
\setlength{\belowdisplayskip}{2pt}



\title{Enhancing UAV Trajectory and Communications Through Vision-Inertial Tracking}




\author{Abdallah~S.~Ghazy\IEEEauthorrefmark{1}, ~\IEEEmembership{Senior Member,~IEEE}, Hussein~A.~Ammar\IEEEauthorrefmark{1},~\IEEEmembership{Member,~IEEE}, James~Bayes\IEEEauthorrefmark{1}, and~Francois~Chan\IEEEauthorrefmark{1},~\IEEEmembership{Senior~Member,~IEEE}
\vspace{-2em}
\thanks{
		\IEEEauthorrefmark{1}Authors are with the Department of Electrical and Computer Engineering (ECE), Royal Military College of Canada, Kingston, ON,  Canada, K7K 7B4.
	}
}

\maketitle

\begin{abstract}
In this paper, we propose an energy-efficient and reliable communication system for non-terrestrial networks deployed in dynamic  GPS-denied wireless environments, enabled by a Vision–Inertial Tracking-Assisted UAV Communication (VIT-UAVCom) system. To the best of our knowledge, this is the first work to exploit onboard UAV cameras and IMU sensors for UAV-assisted communications. We consider a complete VIT-UAVCom system that incorporates  the key design parameters while explicitly accounting for system noise and residual tracking inaccuracies. Building on this framework, we formulate an optimization problem for jointly designing the UAV trajectory and communication performance to improve propulsion energy efficiency, reduce outage probability, and enhance physical-layer security. We then develop a dedicated solution framework to efficiently compute near-optimal trajectory and communication control actions in dynamic scenarios. Furthermore, to enable real-time implementation, we propose and evaluate three optimizers, namely linear search (LS), binary search (BS), and genetic search. Our numerical results demonstrate that our proposed VIT-UAVCom framework significantly outperforms the K-means benchmark in terms of energy consumption while maintaining robust secrecy performance and reliable user coverage. Specifically, our proposed framework improves the energy efficiency by~$144\%$ compared to the benchmark. Interestingly, our results also show that, compared with the LS, the BS reduces the computational time by approximately~$50\%$.
\end{abstract}


\vspace{-1em}

\section{Introduction}
Unmanned aerial vehicles (UAVs) have emerged as a promising enabler of wireless communication networks, serving as flying mobile relays to enhance reliability and capacity. Their inherent mobility allows rapid deployment, dynamic repositioning, and flexible network reconfiguration, making UAVs well suited for temporary networks, traffic offloading in dense areas, emergency communications when terrestrial infrastructure is unavailable or damaged, and non-terrestrial networks (NTNs). Consequently, UAV-assisted wireless systems have attracted significant attention as a versatile solution to modern communication challenges, \cite{Hussein_2026}.

Extensive research has investigated UAV-assisted relay networks, with emphasis on optimal placement, trajectory design, and performance optimization. Early studies demonstrated that optimizing UAV altitude and horizontal positioning can significantly improve achievable rates~\cite{Rate_Optimization}. Mission-critical scenarios were addressed in~\cite{URLLC_Enabled}, where joint UAV placement and finite blocklength design enabled reliable low-latency relaying. Advanced access techniques were explored in~\cite{HS_UAV_NOMA}, revealing strong coupling between NOMA decoding order and UAV positioning. Three-dimensional placement optimization was shown to enhance SNR and link robustness in~\cite{2020_UAV_Relay_3D}.  Interference-aware multi-hop relay deployment was investigated in~\cite{Multi_Hop_UAV_Relay}, and energy-efficient UAV-NOMA relaying was studied in~\cite{UAV_NOMA}. For failure recovery scenarios, model-free reinforcement learning was adopted in~\cite{UAV_Malfunctioning}, and low-complexity local search placement algorithms were proposed in~\cite{Local_Map}. Collectively, these works highlight the central role of intelligent UAV placement and mobility control in improving reliability, coverage, and energy efficiency.

\begin{table*}[t]
	\centering
	\caption{ Literature Survey   of UAV Trajectory Optimization in Dynamic Networks}
    \vspace{-0.5em}
	\label{tab:literature_survey}
	\begin{tabular} {|p{.5cm}|p{.5cm}|p{4cm}|p{6cm}| p{5cm}|}
		\hline
		\textbf{Ref.} & \textbf{Year} & \textbf{Objective}  & \textbf{Contributions}  & \textbf{Localization Technique}   \\
	\hline	
			\cite{2019_Direction_Displacement} & 2019& 
            Optimizing UAV’s displacement distance to maximize the average throughput and the successful transmission probability. & Derive analytic formulas for the optimal displacement for different objectives  & The UAV does not need to learn users’ exact locations in real time, but flying to the spatial sector with the greatest number of users in the cell. \\

				\hline	
				\cite{2020_Solar_UAV} & 2020&Optimizing trajectories for solar-powered UAVs, jointly maximizing communication performance and energy harvesting,  enable efficient UAVs in dynamic environments. & Adopt a distributed model predictive control (DMPC) framework that optimizes UAV control inputs while predicting system states over a receding time horizon.  & UAV positioning and motion are implicitly handled through system state prediction within the predictive control framework, assuming that the required state information is available. \\		
		\hline
		
		\cite{A_New_Framework} & 2022 & 
        Extending the coverage and maximizing the number of served users under backhaul constraints. &  Predict the spatio-temporal distribution of mobile users to adjust the UAVs positions accordingly. & Recurrent Neural Networks (RNN) \\
		\hline
		
		\cite{Adaptive_3D_Placement} & 2023 &  
        Improving long-term mean opinion score compared to rate-maximization baselines. &  Movement design enabling UAV to adjust their 3D
		locations with respect to the real-time network condition. & K-means algorithm   \\
		\hline
		
		\cite{Federated_Learning_Based} & 2024 & 
        Optimizing trajectory of UAV to minimize the outage probability & Deploy UAVs in a proactive, users’ demands-aware, and distributed manner to prevent outage regions, which efficiently minimizes the degradation of Quality of Experience (QoE). & Leveraging users's historical mobility data to forecast their future movements within the network   \\
		\hline
		
		\cite{Learning_Deployment} & 2024 & 
        Maximizing the total network throughput while meeting real-time demands. & Formulate a computer vision problem and use a convolutional neural network (CNN) solve it.
        &  The spatial distribution of the users is predicted versus the time using the dataset associated to the users's  area.
		\\ \hline \cellcolor{myGreen1} 
		Our work & \cellcolor{myGreen1} 2026 & \cellcolor{myGreen1}  
		Minimizing UAV total energy consumption while maintaining a secure reliable access for mobile users in a dynamic and GPS-denied environments.
		
		& \cellcolor{myGreen1}  		
Model a VIT-UAVCom system trhough formulating a joint UAV trajectory and adaptive RF beamforming optimization problem. Develop a sub-optimal solution approach, and propose three optimizers to solve it in a real time.					
	& \cellcolor{myGreen1} Localizing the mobile users in real-time using a VIT system empowered by onboard camera and IMU modules makes the system a cost-efficient.\\
		\hline
		
	\end{tabular}
    \vspace{-2em}
\end{table*}

Several works have investigated the deployment of UAVs in dynamic environments. Table~\ref{tab:literature_survey} summarizes recent advances in mobility-aware UAV placement for rebroadcast wireless networks. Early studies such as~\cite{2019_Direction_Displacement} proposed adaptive UAV displacement strategies based on user spatial distribution without requiring exact user localization. The work in~\cite{2020_Solar_UAV} considered solar-powered UAV trajectory optimization by jointly accounting for communication performance and energy harvesting using distributed model predictive control. Learning-based trajectory optimization approaches were further explored  in~\cite{A_New_Framework}, where recurrent neural networks were used to predict spatio-temporal user distributions.

Clustering-based deployment strategies such as~\cite{Adaptive_3D_Placement} utilized K-means for adaptive 3D positioning. More recent works, including~\cite{Federated_Learning_Based}, leveraged federated learning and historical mobility data for proactive UAV deployment, and~\cite{Learning_Deployment} employed computer vision and convolutional neural networks to predict user spatial distributions and accelerate deployment decisions. Cellular edge coverage enhancement via mobile UAV relays was demonstrated in~\cite{Multicell_Edge_Coverage}. However, \textit{most existing studies assume known user locations} or rely on indirect mobility prediction rather than explicit real-time sensing-based localization, such as vision-based localization algorithms.

When attached to UAVs, onboard cameras can help reduce beam-training overhead in highly mobile wireless links. Early work in~\cite{ZouACM2022} introduced mmWave beamforming for UAV-to-vehicle communications using visual vehicle detection and tracking to support efficient beam alignment. Later studies demonstrated that deep learning models can directly map RGB images to beam indices for mmWave and THz UAV communications, achieving high beam-selection accuracy without requiring exhaustive beam search~\cite{XuTWC2023,JiangGLOBECOMW2022}. These approaches benefit from the strong correlation between line-of-sight (LoS) aerial channels and the visual visibility of users and surrounding objects. Vision-assisted beam management was subsequently extended to V2X and infrastructure-assisted communication systems~\cite{ReusMunsMSN2021}. In addition, multi-modal frameworks combining vision with GPS or positional information were proposed to improve robustness and reliability~\cite{ZhengArxiv2023}. More recent system-level research investigated camera-assisted UAV beamforming and trajectory optimization~\cite{HuaJIOT2023,VTOPA2025}, while end-to-end vision–channel learning frameworks were developed to further reduce beam-training overhead under mobility and environmental uncertainty~\cite{ZhouELL2025,Arxiv2412VisionBeamforming}.

To the best of our knowledge, no prior work has studied a computer-vision-based unified framework that jointly optimizes UAV trajectory, outage probability, and channel secrecy in real time. To address this gap, we propose a Vision–Inertial Tracking-assisted UAV communication (VIT-UAVCom) system. The proposed VIT-UAVCom system employs a Vision–Inertial Tracking (VIT) technique to localize moving ground users in real time; it then jointly optimizes the UAV trajectory and adaptive RF beamforming to minimize UAV's total energy consumption while ensuring secure and reliable communication. The main contributions of our paper can be summarized as:
\begin{itemize}
\item \textbf{Vision–Inertial Tracking-Assisted UAV Communication (VIT-UAVCom)}: We propose a VIT-assisted UAV communication framework using onboard camera and IMU in GPS-denied dynamic networks. The system model captures key interactions and provides an end-to-end design rather than isolated components.

\item \textbf{Joint UAV Trajectory and Communication Optimization}: We formulate a joint optimization of UAV trajectory and communication parameters to improve energy efficiency, outage probability, and secrecy. We develop efficient real-time solvers based on linear search (LS), binary search (BS), and genetic search (GS).

\item \textbf{Performance and Efficiency Evaluation}: We validate the framework against K-means (KM), achieving up to $144\%$ energy efficiency improvement. Among solvers, BS provides the lowest computational time, reducing runtime by about $50\%$ with a slight efficiency trade-off.
\end{itemize}

The rest of the paper is organized as follows. Section~\ref{Syetm_Model} introduces our system model. Section~\ref{UAV-Based_Communication_System_Modeling} defines the elements of our communication system. Section~\ref{Problem_Formulation_and_Real-Time_Algorithm} formulates our optimization problem and shows how to solve it. Section~\ref{Numerical_Results} presents the numerical results. Section~\ref{Conclusion} concludes our work and shows future directions.

\begin{figure*}
	\centering
	\includegraphics[width = .9999\linewidth]{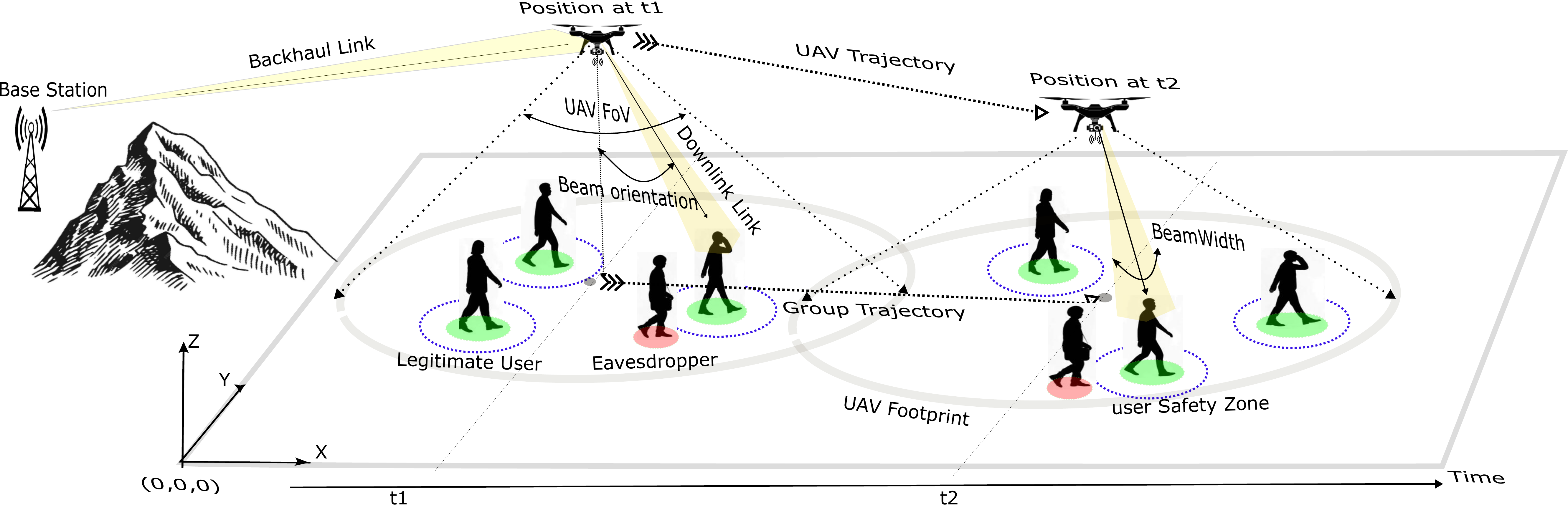}
    \vspace{-1em}
	\caption{Considered network topology}
	\label{Network_Topology}
    \vspace{-1em}
\end{figure*}

\vspace{-0.2em}
\section{  VIT-UAVCom System Model}
\label{Syetm_Model}

\subsection{Overview of the Proposed Framework}
\label{Paper_Framework}

We consider the scenario of providing wireless connectivity to remote users through a UAV-mounted radio unit, as shown in Fig.~\ref{Network_Topology}, where our wireless communication system faces several critical challenges. First, obstructions due to terrain and long distances block direct communication links between the terrestrial base station (BS) and ground users, significantly affecting network connectivity. Second, the absence of GPS services or intentional denial through jamming prevents conventional UAV localization and navigation. Third, the presence of potential eavesdroppers attempting to intercept legitimate transmissions raises physical-layer security concerns.

The UAV uses directional beam steering to serve the users. However, wind-induced turbulence can perturb the UAV trajectory and orientation, resulting in unintended positional deviations and attitude drift. These impairments lead to RF beam misalignment, degradation in beamforming accuracy, and increased signal leakage toward potential eavesdroppers. To overcome these challenges, we propose a VIT-UAVCom system, where the UAV restores wireless connectivity by establishing a directional RF link with the BS (BS-to-UAV link) while simultaneously operating as a flying, tracking ``access point'' that delivers downlink (UAV-to-user link) communication services to a group of mobile ground users.


In the absence of GPS, the VIT module enables real-time user localization (tracking), allowing the UAV to follow a time-varying trajectory from $t_1$ to $t_2$ in order to track group mobility. To ensure mission completion under limited onboard energy resources, the UAV minimizes propulsion energy consumption through trajectory optimization.

To guarantee secure and reliable communication, the UAV adaptively adjusts its RF beam orientation and beamwidth, illustrated by the light-yellow cones in Fig.~\ref{Network_Topology}. This adaptive beamforming strategy concentrates the transmitted energy toward legitimate users (shown as green circles) while mitigating signal leakage toward potential eavesdroppers (shown as light-red circles). Furthermore, a user safety zone is defined around each legitimate user (depicted as dashed-blue circles), representing a protected spatial region assumed to be inaccessible to eavesdroppers. The effective communication coverage area (shown as gray circles) is determined by the UAV footprint, which depends on the UAV altitude, its field of view (FoV), and transmit power.

\begin{algorithm}[t]
	\caption{System Structure}
	\label{Alg_VIT_UAVCom}
	
	\begin{algorithmic}[1]
		
		\State Initialize the VIT-UAVCom system
		
		\While{system is active}
		
		\State Track the users' positions using the VIT system \label{algo:trackUPos}
		
		\State Optimize the UAV trajectory
		\Statex \hspace{0.5cm} - Compute the movement direction
		\Statex \hspace{0.5cm} - Optimize the displacement vector
		
		\State Optimize the RF beam parameters \label{algo:optBeam}
		\Statex \hspace{0.5cm} - Adjust beamwidth
		\Statex \hspace{0.5cm} - Adjust beam orientation
		
		\State Perform time scheduling
		\Statex \hspace{0.5cm} - Schedule the RF beam among users
		
		\State Detect users' motion using the VIT system
		
		\If{user motion is detected}
		\State Go to Step~\ref{algo:trackUPos}
		\Else
		
		\State Collect uplink feedback from users
		
		\If{users are satisfied}
		\State Provide communication service
		\Else
		\State Go to Step~\ref{algo:optBeam}
		\EndIf
		
		\EndIf
		
		\EndWhile
		
	\end{algorithmic}
\end{algorithm}

Algorithm~\ref{Alg_VIT_UAVCom} presents the system structure of our proposed solution. The process starts by tracking the users' positions using the VIT system, which continuously monitors user mobility and network topology variations. Based on the collected information, a motion-detection module determines whether significant user movement has occurred.

\begin{figure*}
	\centering
	\includegraphics[width = .90\linewidth]{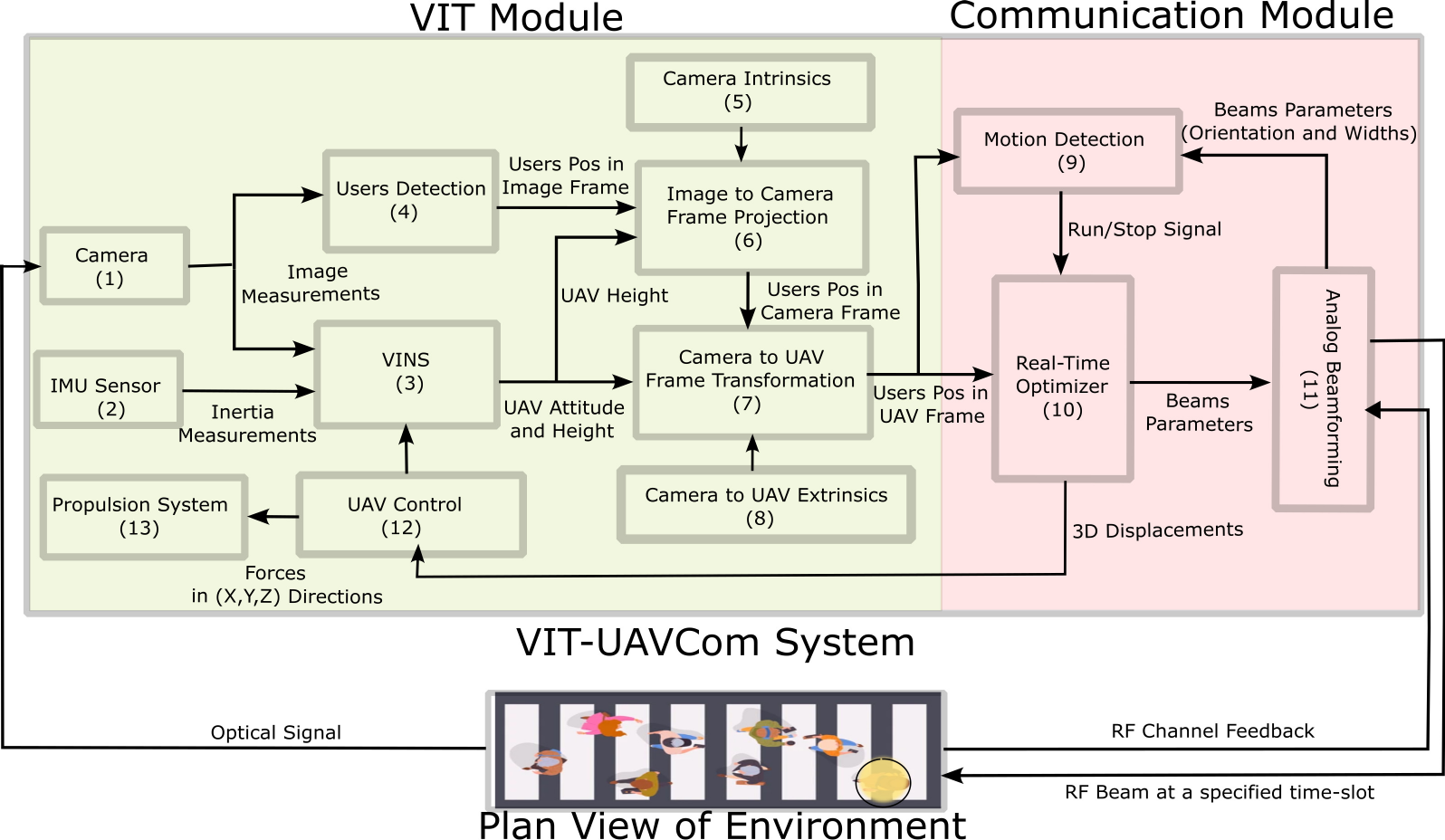}
	\caption{Block Diagram of the VIT-UAVCom System.}
	\label{System_Model}
    \vspace{-1em}
\end{figure*}

If motion is detected, the UAV trajectory optimization block is activated to determine the optimal UAV movement direction and displacement of the users' group to maintain reliable communication. Subsequently, the RF beam parameters, including beamwidth and beam orientation, are optimized to improve communication performance. Afterward, a time-scheduling module allocates the RF beam resources among the active users. 
The system then evaluates whether the users are satisfied based on the received uplink feedback and communication quality metrics. If the users are not satisfied, the optimization loop is re-triggered to further refine the UAV trajectory and RF beamforming parameters. Otherwise, the communication process continues normally. This closed-loop framework enables adaptive UAV trajectory control, dynamic RF beam optimization, and reliable wireless connectivity in dynamic environments.

\vspace{-0.7em}
\subsection{VIT-UAVCom System Model}
\label{VIT_Systems}
\vspace{-0.3em}
Fig.~\ref{System_Model} illustrates the architecture of the proposed VIT-UAVCom system, which consists of two tightly integrated modules: the VIT module and the communication module, highlighted in light yellow and light red, respectively. The VIT module continuously estimates the users' positions relative to the UAV axes in real time (i.e., tracking) using camera and IMU sensors, while the communication module utilizes these estimates to optimize the UAV trajectory and  RF beamforming parameters in real time. In this subsection, we introduce the localization model and its errors.

\subsubsection{Relative Position Estimation}~\\
As shown in Fig.~\ref{System_Model}, the proposed framework enables the UAV to localize users relative to its reference frame through blocks 1-8. The estimated localization information is then utilized to efficiently track users by adjusting the propulsion forces (blocks 9-13) while simultaneously optimizing the RF beamforming parameters (width and orientation) in real time (blocks 9-11).

The VIT module makes use of the camera\footnote{The proposed system requires only a monocular camera, rather than a stereo camera, thereby reducing cost.} (block 1), to capture the mobile users within the camera field-of-view (FoV) angle, $\theta_{\rm FoV}$. To ensure that the camera continuously captures the mobile users, the UAV altitude, $h_u$, should be selected such that the camera footprint radius, $R_{\rm FoV}$, fully covers the maximum spatial extent of the user group, $R_{\max}$, with an additional safety margin, $\Delta$, to account for localization uncertainty and abrupt user motion. Assuming a conical camera projection model and $R_{\rm FoV} > R_{\max} + \Delta$, the camera footprint radius can be approximated as $R_{\rm FoV} = h_u\tan\!\left( \frac{\theta_{\rm FoV}}{2} \right)$. Accordingly, the minimum UAV altitude constraint can be expressed~as
\begin{equation}
h_{u}
>
\frac{
	R_{\max}+\Delta
}{
\tan\!\left(
\theta_{\rm FoV}/2
\right)
},
\label{eq:min_altitude}
\end{equation}
where \eqref{eq:min_altitude} highlights the trade-off between UAV altitude, camera FoV, sensing reliability, and user mobility.

The camera (block 1) produces image measurements of the users. Let $\mathbf{p}_k^w \in \mathbb{R}^{3\times1}$ denote the 3D position of a user in the world frame for $k^{th}$ user. The corresponding image measurement vector $\mathbf{z}_k\in\mathbb{R}^{2\times1}$, for $k^{th}$ user, can be modeled using the pinhole camera model as~\cite[Eq.~23]{Eq_23}
\begin{equation}
\mathbf{z}_k =
\pi \!\left(
\mathbf{R}_{cw}
(\mathbf{p}_k^w-\mathbf{p}_c^w)
\right)
+\mathbf{n}_v,
\label{eq:image_projection}
\end{equation}
where $\pi(\cdot):\mathbb{R}^{3}\rightarrow\mathbb{R}^{2}$ is the nonlinear projection function, $\mathbf{R}_{cw}\in\mathbb{R}^{3\times3}$ denotes the world-to-camera rotation matrix, $\mathbf{p}_c^w \in \mathbb{R}^{3\times1}$ represents the camera position in the world frame, and $\mathbf{n}_v\in\mathbb{R}^{2\times1}$ denotes the  measurement noise.

Simultaneously, the IMU sensor (block 2) measures the UAV inertial motion through linear acceleration and angular velocity measurements. The linear acceleration is measured as~\cite{2019_Calibration}
\begin{equation}
\mathbf{a}_m=
\mathbf{R}_{wb}^{-1}
(\mathbf{a}^w-\mathbf{g})
+\mathbf{b}_a+\mathbf{n}_a,
\label{eq:accelermeter}
\end{equation}
where $\mathbf{a}_m\in\mathbb{R}^{3\times1}$ denotes the measured acceleration, $\mathbf{R}_{wb}\in\mathbb{R}^{3\times3}$ is the body-to-world rotation matrix, $\mathbf{a}^w,\mathbf{g},\mathbf{b}_a,\mathbf{n}_a\in\mathbb{R}^{3\times1}$ denote  world-frame acceleration, gravity vector, accelerometer bias, and sensor noise, respectively. The angular velocity is measured as~\cite{2019_Calibration}
\begin{equation}
\boldsymbol{\omega}_m=
\boldsymbol{\omega}^b
+\mathbf{b}_{\omega}
+\mathbf{n}_{\omega},
\label{eq:gyroscope}
\end{equation}
where $\boldsymbol{\omega}_m,\boldsymbol{\omega}^b,\mathbf{b}_{\omega},\mathbf{n}_{\omega}\in\mathbb{R}^{3\times1}$ denote the measured angular velocity, body-frame angular velocity, gyroscope bias, and sensor noise, respectively.

The Visual Inertial Navigation System (VINS), in block 3, fuses the image measurements from the camera (block 1) and the inertial measurements from the IMU (block 2) to estimate the UAV state matrix as~\cite[Eq.~1, Eq.~2]{Eq_IMU}
\begin{equation}
\mathbf{X}^w_u=
\begin{bmatrix}
\mathbf{p}_u^w &
\mathbf{v}_u^w &
\mathbf{q}_u^{w}
\end{bmatrix}^{T},
\label{eq:vins_state}
\end{equation}
where $\mathbf{X}^w_u\in\mathbb{R}^{3\times3}$ denotes the matrix state of the UAV relative to the world frame,    $\mathbf{p}_u^w,\mathbf{v}_u^w\in\mathbb{R}^{3\times1}$ denote the UAV position and velocity vectors, respectively, while $\mathbf{q}^w_{u}\in\mathbb{R}^{3\times1}$ denotes the UAV attitude vector.

Next, the user detection module (block 4) extracts user bounding boxes and visual feature points from the received image frames using a real-time object detection algorithm, such as YOLO~\cite{Redmon2018YOLOv3}. Using the UAV height information from block 3 and the camera intrinsics from block 5, the detected image coordinates are back-projected through the image-to-camera-frame projection module (block 6) to estimate the users' positions in the camera coordinate frame as~\cite[Eq.~1]{2021_Frame_Transformation}
\begin{equation}
\hat{\mathbf{p}}_k^{c}
=
\lambda\,
\mathbf{K}^{-1}
\tilde{\mathbf{z}}_k,
\label{eq:camera_projection}
\end{equation}
where $\hat{\mathbf{p}}_k^{c}\in\mathbb{R}^{3\times1}$ denotes the estimated user position in the camera frame, $\mathbf{K}\in\mathbb{R}^{3\times3}$ is the camera intrinsic matrix, $\tilde{\mathbf{z}}_k\in\mathbb{R}^{3\times1}$ denotes the homogeneous image coordinate, and $\lambda\in\mathbb{R}^{1\times1}$ is the estimated depth parameter. The calibrated camera intrinsic matrix is defined as~\cite{2023_Camera_Intrinsics}
\begin{equation}
\mathbf{K} =
\begin{bmatrix}
f_x & 0   & c_x \\
0   & f_y & c_y \\
0   & 0   & 1
\end{bmatrix},
\end{equation}
where $f_x$ and $f_y$ denote the focal lengths in pixel units along the horizontal and vertical image axes, respectively, while $(c_x,c_y)$ denotes the principal point.

To estimate the user locations relative to the UAV frame, the camera-to-UAV-frame transformation module (block 7) utilizes the UAV attitude and height information from block 3, together with the camera-to-UAV extrinsic from block 8, to transform the estimated user positions into the UAV reference frame according to~\cite[Eq.~32]{Eq_IMU} and~\cite{Vitual_Inertia_Tracking_2018}
\begin{equation}
\hat{\mathbf{p}}_k^{u}
=
\mathbf{R}_{uc}\,
\hat{\mathbf{p}}_k^{c}
+
\mathbf{t}_{uc},
\label{eq:extrinsic_transform}
\end{equation}
where $\hat{\mathbf{p}}_k^{u}\in\mathbb{R}^{3\times1}$ denotes  estimated user position in the UAV frame, $\mathbf{R}_{uc}\in\mathbb{R}^{3\times3}$ denotes the camera-to-UAV rotation matrix, and  $\mathbf{t}_{uc}$ denotes the translation vector. At this stage, the UAV has fully estimated the user's locations in its reference frame.

The communication module then utilizes the estimated users’ positions and incorporates a motion detection module (block 9) to estimate user displacement. Typically, block 9 compares the current RF beam footprints from block 11 with the current estimated user positions. If significant mismatches are detected, this indicates that the current RF beam configuration is no longer optimized. Consequently, block 9 triggers the real-time optimizer (block 10) to execute the optimization process. The real-time optimizer jointly determines the optimized UAV displacement vector and beamforming parameters.

The UAV trajectory controller (block 12) generates a motion-control signal for the propulsion system (block 13) according to the optimized displacement vector. The propulsion force vector is modeled as~\cite[Eq.~3]{Force_Equation}
\begin{equation}
\mathbf{f}_{p} =
m\,\ddot{\mathbf{p}}
+
m\,\mathbf{g}
-
\mathbf{f}_{w},
\label{eq:uav_dynamics}
\end{equation}
where $\mathbf{f}_{p},\ddot{\mathbf{p}}, \mathbf{g},  \mathbf{f}_{w} \in\mathbb{R}^{3\times1}$ denote the propulsion force vector, UAV acceleration vector, the gravitational acceleration vector, and wind-induced force vector,  respectively. Moreover, $m\in\mathbb{R}^{1\times1}$ denotes the UAV mass. Likewise, block 12 feeds its updates back to block 3, facilitating the UAV attitude and height estimation process.

Once receiving inputs from block 10, the adaptive analog beamforming module (block 11) emits directional RF beams toward the legitimate users by dynamically controlling the beam orientation and beamwidth for each user at every time slot. Let $\mathbf{w}[n]\in\mathbb{C}^{N_q\times1}$ denotes the analog beamforming vector at time slot $n$, and $N_q$ denotes the number of transmit antennas at the UAV. The transmitted beamformed signal can be expressed as
\begin{equation}
\mathbf{x}[n]
=
\mathbf{w}[n] s[n]+\mathbf{n}[n],
\label{eq:beamforming_signal}
\end{equation}
where $\mathbf{x}[n]\in\mathbb{C}^{N_q\times1}$ denote the transmitted signal vector, while $s[n]\in\mathbb{C}^{1\times1}$ and $\mathbf{n}[n]\in\mathbb{C}^{N_q\times1}$ represent the transmitted communication symbol and system noise, respectively.

Overall, the proposed VIT-UAVCom system establishes a continuous closed-loop interaction between the VIT module and the communication module, enabling the UAV to efficiently track legitimate users while providing reliable and secure communication services in real time, independently of GPS availability.

\subsubsection{Residual Localization Errors}~\\
\label{subsec:relative_localization_uncertainty}
This subsection models the residual localization errors of the estimates discussed in the previous subsection. In contrast to Simultaneous Localization and Mapping (SLAM)-based navigation methods, where errors typically accumulate over time due to long-horizon state propagation, our proposed VIT framework performs instantaneous vision-aided localization updates. As a result, the localization error remains bounded and does not exhibit unbounded drift. However, short-term temporal correlation may still arise due to slowly varying IMU biases~\cite{Eq_IMU}.

The error model relies on the following assumptions: (i) users are located on a locally planar surface; (ii) a near-nadir camera geometry is considered, such that horizontal errors dominate; (iii) camera intrinsics and camera-to-UAV extrinsics are fixed and pre-calibrated; (iv) UAV altitude and attitude are obtained from inertial sensing; and (v) the sensing model is locally linear, enabling first-order uncertainty propagation. Under these assumptions, the relative localization error for the $k^{th}$ user is defined as
\begin{equation}
\tilde{\mathbf{p}}^{u}_k \triangleq \mathbf{p}^{u}_k - \hat{\mathbf{p}}^{u}_k,
\label{Eq_Error}
\end{equation}
where $\tilde{\mathbf{p}}^{u}_k, \hat{\mathbf{p}}^{u}_k \in \mathbb{R}^{3\times1}$ denote the residual error and estimated user position vectors relative to the UAV frame, respectively. The residual error is modeled as a zero-mean Gaussian random vector~\cite{2004_Imaging_TextBook}
\begin{equation}
\tilde{\mathbf{p}}^{u}_k \sim \mathcal{N}\!\big(\mathbf{0},\boldsymbol{\Sigma}_u\big),
\end{equation}
where $\boldsymbol{\Sigma}_u \in \mathbb{R}^{3\times3}$ is the covariance matrix capturing visual, inertial, and calibration uncertainties.

Linearizing the nonlinear geometric mapping yields the first-order covariance propagation~\cite{Thrun2005Probabilistic}
\begin{equation}
\boldsymbol{\Sigma}_u =
\mathbf{J}_{\text{cam}}\boldsymbol{\Sigma}_{\text{cam}}\mathbf{J}_{\text{cam}}^{\mathsf T}
+
\mathbf{J}_{\text{imu}}\boldsymbol{\Sigma}_{\text{imu}}\mathbf{J}_{\text{imu}}^{\mathsf T}
+
\mathbf{J}_{\text{cal}}\boldsymbol{\Sigma}_{\text{cal}}\mathbf{J}_{\text{cal}}^{\mathsf T},
\end{equation}
where $\mathbf{J}_{x}$ and $\boldsymbol{\Sigma}_{x}$ denote the Jacobian and covariance associated with $x \in \{\text{cam},\text{imu},\text{cal}\}$, \cite{2020_Image_Noises}, \cite{2018_IMU}, \cite{2019_Calibration}. Specifically,
$\mathbf{J}_{x} \in \mathbb{R}^{3\times m_x}$ and $\boldsymbol{\Sigma}_{x} \in \mathbb{R}^{m_x\times m_x}$, where $m_x$ denotes the dimension of the corresponding noise source.

Under symmetric sensing conditions, the covariance matrix  is approximated as isotropic,
\begin{equation}
\boldsymbol{\Sigma}_u \approx \sigma_u^2 \mathbf{I}_3,
\end{equation}
where $\mathbf{I}_3 \in \mathbb{R}^{3\times3}$ and
\begin{equation}
\sigma_u^2 =
\sigma_{\text{cam}}^2
+
\sigma_{\text{imu}}^2
+
\sigma_{\text{cal}}^2.
\end{equation}

Using first-order uncertainty propagation of the pinhole camera model~\cite{Triggs2000BA}, the camera-induced horizontal localization variance under near-nadir geometry can be approximated as
\begin{equation}
\sigma_{\text{cam}}^2 =
h_u^2
\left(
\frac{\sigma_{\text{pix}}^2 + \sigma_{\text{det}}^2}{2}
\left(
\frac{1}{f_x^2} + \frac{1}{f_y^2}
\right)
\right),
\end{equation}
where $\sigma_{\text{pix}},\sigma_{\text{det}}$ denote pixel noise and detection uncertainty.

Inertial effects are captured using an effective variance model, $\sigma_{\text{imu}}^2$, which accounts for IMU-driven error accumulation between consecutive camera updates $T_c$, rather than long-term drift. Following standard IMU stochastic error propagation models~\cite{Titterton2004INS}, the effective IMU-induced localization variance is approximated as
\begin{equation}
\sigma_{\text{imu}}^2 =
k_a\,\sigma_a^2\, T_c^3
+
k_{ab}\,\sigma_{ab}^2\, T_c^5
+
k_w\, g^2\,\sigma_w^2\, T_c
+
k_{wb}\, g^2\,\sigma_{wb}^2\, T_c^3,
\label{Eq_Error_IMU}
\end{equation}
where $\sigma_a$ and $\sigma_w$ are the accelerometer and gyroscope white-noise densities, $\sigma_{ab}$ and $\sigma_{wb}$ are the corresponding bias random-walk coefficients \cite{2018_IMU_Biase}, $T_c$ is the camera update period, and $g$ denotes gravitational acceleration.  The coefficients $k_a$, $k_{ab}$, $k_w$, and $k_{wb}$ are positive, dimensionless constants that capture motion- and geometry-dependent scaling effects within the VIT system error-state EKF. 

To model harsher sensing conditions, the covariance may be inflated as \cite{Maybeck1979Stochastic}
\begin{equation}
\tilde{\mathbf{p}}^{u}_k \sim \mathcal{N}\!\big(\mathbf{0},\alpha \boldsymbol{\Sigma}_u\big),
\end{equation}
where $ \alpha \ge 1$.

To capture temporal correlation due to slowly varying IMU biases, the error is optionally modeled as a first-order Gauss--Markov process \cite{Maybeck1979Stochastic}
\begin{equation}
\tilde{\mathbf{p}}^{u}_k[n] =
\rho\,\tilde{\mathbf{p}}^{u}_k[n-1] + \mathbf{n}_p[n],
\qquad 0 \le \rho < 1,
\end{equation}
where $\mathbf{n}_p[n] \in \mathbb{R}^{3\times1}$. The noise process satisfies \cite{Maybeck1979Stochastic}
\begin{equation}
\mathbf{n}_p[n] \sim \mathcal{N}\!\big(\mathbf{0}, (1-\rho^2)\boldsymbol{\Sigma}_u\big),
\end{equation}
with correlation coefficient \cite{Hesch2014Observability}
\begin{equation}
\rho = \exp\!\left(-\frac{\Delta t}{\tau}\right),
\label{Taue}
\end{equation}
where $\Delta t$ is the sampling interval and $\tau$ is the correlation time constant.

Table~\ref{tab:noise_levels} summarizes the parameters of the localization error model, Eqs.~\eqref{Eq_Error}-\eqref{Taue}, under low-, medium-, and high-noise conditions, consistent with established camera, IMU, and vision-based detection models~\cite{2004_Imaging_TextBook,Scaramuzza2011VisualOdometry,Redmon2018YOLOv3,Triggs2000BA,Titterton2004INS,Grewal2013INS,Hesch2014Observability}. The calibration noise assumes small camera-to-UAV extrinsic errors, corresponding to rotational misalignments of $0.1^\circ$-$1^\circ$ and translational offsets of $1$-$5$~cm, projected onto the ground plane at $h_u=100$~m.

\begin{table}[t]
	\centering
	\caption{Noise parameters of the residual error model, \cite{2024_Simulation_Table, 2004_Imaging_TextBook,Scaramuzza2011VisualOdometry,Redmon2018YOLOv3,Triggs2000BA,Titterton2004INS,Grewal2013INS,Hesch2014Observability}.}
	\label{tab:noise_levels}
	\setlength{\tabcolsep}{4pt}
	\renewcommand{\arraystretch}{1.5}
	\begin{tabular}{lccc c}
		\hline
		\textbf{Parameter} & \textbf{Low} & \textbf{Medium} & \textbf{High} & \textbf{Ref.} \\
		\hline
		\multicolumn{5}{c}{\emph{Camera and Visual Front-End}} \\
		\hline
		$f_x,f_y$ (pixel) & 600 & 500 & 400 & \cite{2004_Imaging_TextBook} \\
		$\sigma_{\text{pix}}$ (pixel) & 0.5 & 1.0 & 2.0 & \cite{2004_Imaging_TextBook} \\
		$\sigma_{\text{det}}$ (m) & 0.05 & 0.20 & 0.50 & \cite{Redmon2018YOLOv3} \\
			$\alpha$ (dimensionless) & 1 & 0.9 & 0.5 & -- \\
		\hline
		\multicolumn{5}{c}{\emph{IMU Parameters}} \\
		\hline
		$\sigma_a$ (m/s$^2/\sqrt{\text{Hz}}$) & $5\!\times\!10^{-3}$ & $2\!\times\!10^{-2}$ & $5\!\times\!10^{-2}$ & \cite{Titterton2004INS,Grewal2013INS} \\
		$\sigma_w$ (rad/s$/\sqrt{\text{Hz}}$) & $1\!\times\!10^{-3}$ & $2\!\times\!10^{-3}$ & $5\!\times\!10^{-3}$ & \cite{Titterton2004INS,Grewal2013INS} \\
		$\sigma_{ab}$ (m/s$^3/\sqrt{\text{Hz}}$) & $5\!\times\!10^{-4}$ & $1\!\times\!10^{-3}$ & $3\!\times\!10^{-3}$ & \cite{Hesch2014Observability} \\
		$\sigma_{wb}$ (rad/s$^2/\sqrt{\text{Hz}}$) & $2\!\times\!10^{-5}$ & $5\!\times\!10^{-5}$ & $1\!\times\!10^{-4}$ & \cite{Hesch2014Observability} \\
		$\tau$ (seconds) & 5 & 10 & 20 & --\\
		\hline
		\multicolumn{5}{c}{\emph{Derived Quantities (assuming $h=100$ m and $T_c=1/30$ s)}} \\
		\hline
		$\sigma_{\text{cam}}$ (m) & 0.08 & 0.20 & 0.50 & \cite{2004_Imaging_TextBook} \\
		$\sigma_{\text{imu}}$ (m) & 0.30 & 1.00 & 2--3 & \cite{Titterton2004INS,Hesch2014Observability} \\
		$\sigma_{\text{cal}}$ (m) & 0.05 & 0.15 & 0.40 & -- \\
$\sigma_u$ (m) & 0.33 & 1.06 & 2.2--3.3 & -- \\
		\hline
	\end{tabular}
    \vspace{-2em}
\end{table}

\vspace{-1em}
\section{Communication System Modeling}
\label{UAV-Based_Communication_System_Modeling}
We consider a terrestrial BS that provides a wireless backhaul link to an UAV, where the UAV (Tx) serves $K$ legitimate ground users (Rxs) over the downlink, while $K$ potential eavesdroppers (Exs) attempt to intercept the transmitted signals. In this section, we model the downlink (from the BS to the UAV and from the UAV to the users) and evaluate its communication performance.

\vspace{-1em}
\subsection{Network Mobility Model}
Both the UAV and users are mobile, resulting in time-varying channels and dynamic coverage requirements. To capture realistic user behavior, we adopt a \emph{point-reference group mobility model (PRGM)}, \cite{Simulation_Parameters_1}, in which users move around a common, time-varying group center denoted by $\mathbf{c}[n]\in\mathbb{R}^{2\times 1}$ (i.e., reference point)\footnote{The reference point may follow a predefined or adaptive trajectory reflecting coordinated team motion, commonly used in search-and-rescue, disaster relief, scientific, military, and exploration missions in extreme environments such as disaster zones and the Arctic.}.   The position of user $k$, relative to the world frame,  at time slot $n$ is modeled as
\begin{equation}
\mathbf{p}^w_k[n] = \mathbf{c}[n] + r_k[n]
\begin{bmatrix}
\cos\vartheta_k[n]\\
\sin\vartheta_k[n]
\end{bmatrix},
\qquad k=\{1,\ldots,K\}\,,
\label{eq:PRGM_user_position}
\end{equation}
where users are assumed to be uniformly distributed within a disk of radius $R[n]$, yielding $r_k[n] \sim \mathcal{U}\!\left(0,R[n]\right), 
\vartheta_k[n]\sim \mathcal{U}(0,2\pi)$. The group radius $R[n]$ evolves over time to capture expansion and contraction of the user cluster during different mission phases.
A bounded evolution model is adopted as
\begin{equation}
R[n] =
\mathrm{clip}\!\left(
R[n-1] + \nu[n],\,
R_{\min},\,
R_{\max}
\right),
\label{eq:PRGM_radius}
\end{equation}
where the $\mathrm{clip}(\cdot)$ function constrains $R[n]$ to remain within the range $[R_{\min}, R_{\max}]$, where $R_{\min}$ and $R_{\max}$ denote the minimum and maximum allowable group radii, respectively. Specifically, values smaller than $R_{\min}$ are set to $R_{\min}$, while values larger than $R_{\max}$ are set to $R_{\max}$. Moreover, $\nu[n]$ is a zero-mean stochastic or deterministic value specified by users.

\vspace{-1em}

\subsection{Path Loss}
To model the system geometry, we consider the UAV coordinate system as the global coordinate system $(X_o,Y_o,Z_o)$, with the UAV (Tx) located at the origin $O=(0,0,0)$. Relative to this origin $O$,  the BS, Rxs, and Exs are located at $(x_b,y_b,z_b)$, $(x_{k},y_{k},z_k)$,   and $(x_{e},y_{e},z_e)$, respectively.  To enhance physical-layer security, safety zones with radius $d_{s}$ are introduced around each Rx, restricting the proximity of potential eavesdropper. The distance between the $k^{th}$ Rx and its associated Ex is computed as $d_{s}= \sqrt{(x_k-x_e)^2 + (y_k-y_e)^2}$.  The RF unit\footnote{In Canada, UAV wireless communications must comply with spectrum management regulations established by Innovation, Science and Economic Development Canada (ISED). These regulations govern frequency allocation, transmitter certification, and RF emission limits to prevent interference with licensed services. Common UAV communication systems operate in license-exempt bands such as 2.4 GHz and 5.8 GHz, while aviation-protected bands such as 5030–5091 MHz may be used for command and control links in advanced UAV operations.}, attached to the UAV, emits a Gaussian RF beam toward the $k^{th}$ Rx with  transmit power $p_{k}$ and adjustable beamwidth $w_{k}$. 

A strong line-of-sight (LoS) condition is assumed for the BS-to-UAV link, and its path loss is computed as  \cite{2014_PathLoss}
\begin{equation}
g_{b}= 10^{-\Big(20\log_{10}\!\Big(\frac{4\pi f_c d_{b}}{c}\Big)+\eta_{b}^{\mathrm{LoS}}\Big)/10}.
\label{eq:gain_backhaul}
\end{equation}
where $\eta_{b}^{\mathrm{LoS}}$ is environment-dependent excessive losses. For a  predominantly  LoS, $\eta_{b}^{\mathrm{LoS}}$ can be set  to a small value or incorporated into the link-margin term. The symbol $f_c$ denotes the carrier frequency and $c$ is the speed of light, and $d_{b}$ is the BS-UAV distance, given by $d_{b}= \sqrt{x_b^2 + y_b^2 + z_b^2}$. 

The UAV-to-$k^{th}$ Rx downlink (hereafter referred to as the $k^{th}$ link) is characterized by a high LoS probability, which is typical of rural and suburban environments. 
We adopt air-to-ground large-scale channel model,  that accounts for the a LoS probability, LoS and NLoS pathloss, and an average path loss. The corresponding path loss is~\cite{2014_PathLoss}
\begin{equation}
\begin{aligned}
&g^{l}_{k} =\\&10^{-\frac{1}{10} \Bigg( \frac{ \eta_{\mathrm{LoS}}-\eta_{\mathrm{NLoS}}}{1+a\exp\!\left(-b\left(\frac{180}{\pi}\psi_k-a\right)\right)} +20\log_{10}(d_{k})+20\log_{10}\!\left(\frac{4\pi f_c}{c}\right)+\eta_{\mathrm{NLoS}}\Bigg)}
\end{aligned}
\label{eq:gain_access}
\end{equation}
where $\eta_{\mathrm{LoS}}$ and $\eta_{\mathrm{NLoS}}$ are environment-dependent excessive losses, $a$ and $b$ are environment-dependent  empirical parameters,  $d_{k}$ is the  distance between UAV and $k^{th}$ user, and it is computed as $d_{k}= \sqrt{x_k^2 + y_k^2 + z_k^2}$. The $\psi_k$ is the corresponding elevation angle, and it is obtained as $ \psi_k= \sin^{-1}\!\Big(\frac{z_k}{d_{k}}\Big)$.

\vspace{-1em}
\subsection{Wind-Induced Pointing Errors}

In practical operation, wind-induced turbulence causes random UAV displacement and orientation fluctuations, leading to pointing errors and performance degradation. Since the BS-to-UAV link employs a wide RF beam and no nearby eavesdroppers are assumed, pointing errors on this link are neglected. In contrast, the UAV-to-user link is highly sensitive to beam misalignment because the UAV must carefully control its beamwidth to maintain energy efficiency and limit information leakage toward potential eavesdroppers. To model wind-induced fluctuations, we adopt the experimentally validated statistical models in \cite{Ghazy_UAV_PDFs}, which are based on hovering-drone measurements reported in \cite{2013_Hovering_PDF}. The UAV displacement along the $X_o$, $Y_o$, and $Z_o$ axes is modeled by independent Gaussian random variables \cite{2013_Hovering_PDF},
$x_t\sim \mathcal{N}(0,\sigma_{x}),\, y_t\sim \mathcal{N}(0,\sigma_{y}),\, z_t\sim \mathcal{N}(0,\sigma_{z})\,$,
where $\sigma_x$, $\sigma_y$, and $\sigma_z$ denote the corresponding standard deviations. Experimental observations indicate that lateral fluctuations dominate the vertical component, i.e., $\sigma_x=\sigma_y \gg \sigma_z$ \cite{Ghazy_TAMIMO}. The nominal beam axis toward user $k$ is denoted by $\mathcal{Z}k$. Wind-induced disorientation tilts this axis to a new orientation, $\hat{\mathcal{Z}}k$, characterized by the polar and azimuth angles $(\hat{\theta}k,\hat{\phi}k)$. Since UAV orientation fluctuations have a more pronounced effect than receiver dynamics, only UAV disorientation is considered. Following \cite{Cox_Munk}, the polar angle is modeled using the Cox--Munk PDF,
\begin{equation}
{\rm CM}(\hat{\theta_k};\sigma{\hat{\theta}}):=
\dfrac{\tan(\hat{\theta}k)\sec^2(\hat{\theta}k)}
{\pi\sigma{\hat{\theta}}^2}
\exp\left(
-\dfrac{\tan^2(\hat{\theta}k)}
{2\sigma{\hat{\theta}}^2}
\right),
\end{equation}
while the azimuth angle follows the truncated Gaussian PDF \cite{2013_Hovering_PDF}
\begin{equation} {\rm TG}(\hat{\phi_{k}};\mu_{\hat{\phi}},\sigma_{\hat{\phi}}) := \begin{cases} \frac{ \exp\!\left( -\dfrac{(\hat{\phi_{k}}-\mu_{\hat{\phi}})^2}{2\sigma_{\hat{\phi}}^2} \right) }{ \sigma_{\phi}\sqrt{2\pi} \left( 1-\Phi\!\left(-\dfrac{\mu_{\hat{\phi}}}{\sigma_{\hat{\phi}}}\right) \right) }, & \phi_{k} \ge 0, \\[12pt] 0, & \phi_{k} < 0\,, \end{cases} \end{equation}
where $\mu_{\hat{\phi}}$ and $\sigma_{\hat{\phi}}$ are the mean and standard deviation of $\hat{\phi}_k$, respectively, and $\Phi(\cdot)$ denotes the standard normal CDF. The PDF is truncated to satisfy $\hat{\phi}_k \ge 0$. The PDF of the pointing error losses between the UAV and $k^{th}$ user, $g^{p}_{k}$, is given as \cite{Farid_Steve} \begin{equation} f_{g}(g^{p}_{k})= \dfrac{{\xi}^2_{k}}{ A^{{\xi}^2_{k}}} {g^{p}_{k}}^{({\xi}^2_{k}-1)}\,,\quad 0\leq g^{p}_{k} \leq {A}_{k}\,, \label{PDF_Pointing_Error} \end{equation} where ${\xi}^2_{k}=\dfrac{\sqrt{\pi}\, \erf(\epsilon_{k}) \,w_{k}^2}{8\, \Lambda_{k}^2\, \epsilon_{k}\, \exp(-\epsilon_{k}^2)}$, $A_{k}=\left[\erf\left(\epsilon_{k}\right)\right]^2$, and $\epsilon_{k}=\dfrac{\sqrt{\pi}\,a_{k}}{\sqrt{2}\,w_{k}}$, where $\Lambda_{k}^2$, $a_{k}$ and $w_{k}$ are the jitter variance, aperture radius, and the beamwidth of $k^{th}$ Rx, respectively. ~\eqref{PDF_Pointing_Error} provides a good approximation when the beam width is significantly larger than the receiving aperture, i.e., $w_{k}> 6\,a_{k}$~\cite{Farid_Steve}.

\vspace{-0.5em}
\subsection{Channel Secrecy and Outage Probability}
For $k^{th}$ link, the signal-to-noise ratio  is given by
\begin{equation}
\gamma_{k}=\frac{g_{u}\; g^p_{k}\; g^{l}_{k}\;p_{k}}{\sigma_k^2},
\end{equation}
where $g_{u}$ is UAV's antenna gain, and $\sigma_k^2$ is the noise power at the $k^{th}$ Rx. The channel capacity (i.e., maximum achievable bit rate) is given by \cite{2005_Goldsmith_Wireless}
\begin{equation}
c_{k}=B_w\log_2\big(1+\gamma_{k}\big) \leq c_{b} \quad\text{bpcu}\,,
\label{eq:backhaul_rate}
\end{equation}
where $B_w$ is the UAV  bandwidth, $c_{b}$ is the channel capacity  of  the BS-to-UAV link, and $\text{bpcu}$ represents the number of bits per channel use. To guarantee a channel capacity, $c_{k}$, over a bandwidth, $B_w$,
the required average transmitting power can be expressed as
\begin{equation}
	\begin{split}
		&\hat{p}_{k}
		=  g_{u}\; \hat{g}^p_{k}\;
		\sigma_k^2\!\left(2^{\frac{c_{k}}{B_w}}-1\right)
		\times \\ & \hspace{2cm}  10^{\frac{1}{10}\,\left(	\frac{\eta_{\mathrm{LoS}}-\eta_{\mathrm{NLoS}}}
			{1 + a\exp\!\left[-b\frac{180}{\pi}
				\left(\frac{\pi}{2}-\frac{\theta_{\rm FoV}}{2}\right)+ba\right]}\right)} \times \\ &  \hspace{4cm}  10^{\frac{1}{10}\,\left(	20\log_{10}\!\left(\frac{4\pi f_c\,R_s}{c\,\sin\left(\frac{\theta_{\rm FoV}}{2}\right)}\right)
			+ \eta_{\mathrm{NLoS}}\right)}
	\end{split}	
	\label{Power_Threshold}
\end{equation}
where $\hat{g}^p_{k}$ is the average pointing loss of the $k^{th}$ Rx. The outage probability of the $k^{th}$ link, $F_{{k}}(c_{th})$, is calculated as 
\begin{equation}
F_{{k}}(c_{th})=\int_{0}^{c_{th}} f_{{k}}(c_{k})\, dc_{k}\,,
\label{P_out}
\end{equation}
where    $c_{th}$ is channel capacity threshold. Additionally, $f_{{k}}(c_{k})$ is
the PDF of the $k^{th}$ link,  calculated as
\begin{equation}
f_{{k}}(c_{k})= f_{g}(c_{k})  \left|\dfrac{dg^{p}_{k}}{dc_{k}}\right| \,,
\end{equation}
where $f_{g}(c_{k})$ is
the PDF of  DC gain for  $k^{th}$ Rx link. The channel secrecy rate (${\rm CSR}$) is a common metric used in the physical layer security evaluation \cite{Ghazy_UAV_PDFs}. The ${\rm CSR}$ measures the difference between the channel capacity of $k^{th}$ Rx link and that of its associated  Ex. It is defined as
\begin{equation}
{{\rm CSR}_{k}}=\operatorname*{MAX}(0,\hat{c}_{k}-\hat{c}_{{k^{'}}})\,,
\label{CSR}
\end{equation}
where $\hat{c}_{{k}}$ and $\hat{c}_{{k^{'}}}$ are average channel capacities of $k^{th}$ Rx link and ${k^{'}}^{th}$ Ex link, respectively. The average channel capacity of the $k^{th}$ link is computed as 
\begin{equation}
\hat{c}_{k}=\int_{0}^{A_k\,g^l_{k}} c_{k}(g^p_{k})\, \, f_{g}(g^p_{k})\, dg^p_{k}\,\,, \text{bpcu}\,\,,
\label{Average_Capacity}
\end{equation}
as well, the average channel capacity of the ${k^{'}}^{th}$ Ex link is computed as 
\begin{equation}
	\hat{c}_{{k^{'}}}=\int_{0}^{A_{k^{'}}\,g^l_{{k^{'}}}} c_{{k^{'}}}(g^p_{{k^{'}}})\, \, f_{g}(g^p_{{k^{'}}})\, dg^p_{{k^{'}}}\,\, \text{bpcu}\,.
\end{equation}

\subsection{Energy Consumption Model}

We consider both propulsion energy, which is divided into propulsion  and hovering energies, denoted as $E_{\rm pro}$ and $E_{\rm hov}$, respectively, and communication energy, $E_{\rm com}$.  
The total UAV energy consumption, $E_{tot}$, during one mission is given by
\[
E_{tot}=E_{\rm pro}+E_{\rm hov}+E_{\rm com},
\]
and can be expressed as \cite{2019_Pwer_Hovering_Model}

\begin{equation}
\begin{aligned}
&E_{\mathrm{tot}}
=\\&
\underbrace{
	\Bigg(
	P_0\!\left(1+\frac{3V^2}{U_{\mathrm{tip}}^2}\right)
	\!
    +
    \!
	P_i\!\left(\sqrt{1+\frac{V^4}{4v_0^4}}-\frac{V^2}{2v_0^2}\right)^{\!\frac{1}{2}}
	\!\!\!\!
    +
    \!
	\frac{1}{2}\,\hat d_0\, \rho\, s\, A\, V^3
	\Bigg)
}_{\text{Propulsion  Energy } E_{\mathrm{pro}}}
\\[-5pt]
&\hspace{1.2cm} \times	\frac{d_{\mathrm{trav}}}{V} +
\underbrace{
	\left(
	\frac{\delta}{8}\,\rho\, s\, A\, \xi^3\, r^3
	+
	(1+q)\sqrt{\frac{W^3}{2\,\rho\, A}}
	\right)
	T_h
}_{\text{Hovering Energy } E_{\mathrm{hov}}}
\\[-5pt]
&\hspace{3cm} +
\underbrace{
	\sum_{n=1}^{N} p_{k}[n]\,T_s
	+
	P_c\, N\, T_s
}_{\text{Communication Energy } E_{\mathrm{com}}}\,,
\label{eq:total_energy_model}
\end{aligned}
\end{equation}

\noindent
where $\rho$ denotes the air density (kg/m$^3$), 
$\delta$ is the profile drag coefficient, 
$s$ is the rotor solidity, 
$A$ is the rotor disc area (m$^2$), 
$\xi$ is the blade angular velocity (rad/s), 
$r$ is the rotor radius (m), 
$q$ is the incremental correction factor of induced power, 
$W = mg$ is the UAV weight (N) with $m$ being the UAV mass (kg) and $g$ the gravitational acceleration (m/s$^2$), 
$V$ is the UAV traveling speed (m/s), 
$U_{\mathrm{tip}}$ is the rotor blade tip speed (m/s), 
$v_0$ is the mean rotor induced velocity in hover (m/s), 
$\hat d_0$ is the fuselage drag ratio, 
$d_{\mathrm{trav}}$ is the traveling distance (m), 
$T_h$ is the hovering duration (s), 
$N$ is the number of communication slots, 
$T_s$ is the slot duration (s), 
$p_{k}[n]$ is the transmitted power  (W) to $k^{th}$ Rx at slot $n$, 
and $P_c$ is the constant circuit power consumption (W).

\section{Problem Formulation and Solution}
\label{Problem_Formulation_and_Real-Time_Algorithm}
In the system design shown in Fig.~\ref{System_Model}, we introduce a fundamental component referred to as the real-time optimizer (block 10). 
This component is crucial to the proposed VIT-UAVCom system, as it minimizes UAV energy consumption while satisfying network-security and outage-capacity constraints.

In this section, we formulate an optimization problem that balances these objectives and develop a real-time optimizer capable of obtaining a sub-optimal solution within a short processing time.

\vspace{-1em}

\subsection{Problem Formulation}
Our problem formulation aims to minimize the total energy consumption of the UAV, while satisfying minimum secure communication requirements and maintaining the outage probability below a specified threshold. To minimize the total energy consumption $E_{\mathrm{tot}}$, the UAV propulsion energy should be minimized, as it constitutes the dominant component of the overall energy consumption given in ~\eqref{eq:total_energy_model}, i.e., $E_{\mathrm{pro}} > (E_{\mathrm{hov}} + E_{\mathrm{com}})$. Reducing travel energy requires shortening the flight path \footnote{In addition to shortening the flight path, unnecessary stops should be limited, thereby decreasing acceleration and deceleration losses.}, which means that the UAV would spend more time hovering.

To ensure secure communication, the secrecy rate defined in ~\eqref{CSR} must exceed a predefined threshold. This is achieved by focusing the beam coverage on the Rx locations while avoiding coverage of the Ex locations, i.e., by narrowing the RF beam width such that $w_k < w_{\mathrm{th}}$, where $w_{\mathrm{th}}$ is  beamwidth threshold. To guarantee reliable access, Rxs links capacities must satisfy the outage constraint in ~\eqref{P_out}. This typically requires widening the RF beamwidth to improve coverage robustness, i.e., $w_{k} > w_{\mathrm{th}}$.

These objectives are conflicting and introduce fundamental trade-offs. For example, minimizing energy consumption favors shorter UAV trajectories that do not strictly follow user mobility, whereas maximizing secrecy and reliability encourages closer tracking of users to reduce path loss. Moreover, security and reliability are inherently conflicting as widening the beam enhances link reliability but degrades secrecy, and vice versa. Wind-induced fluctuations further intensify these trade-offs. To solve this multi-objective optimization problem, we transform it into a constrained single-objective optimization problem using the $\epsilon$-constraint method.

\begin{figure*}[h]
	\stepcounter{equation}  
	\begin{subequations}
		\begin{align}
			&\underset{\{\mathbf{p}_u[n]|_{n=1}^{N_t},{w}_{k}[n]|_{n=1}^{\sum_{k=1}^{K}N_{b,k}}\}}{\text{minimize}}\,\,
			E_{\mathrm{tot}} (T)
			= \sum_{n=1}^{N_t}
			\left( 
			E_{\mathrm{pro}}\big(\mathbf{p}_u[n],\mathbf{p}_u[n\!-\!1],\dot{\mathbf{p}}_u[n]\big) 
			+ E_{\mathrm{hov}}\big(\mathbf{p}_u[n]\big)\right) 
			+\sum_{k=1}^{K} \sum_{n=1}^{N_{b,k}} \dfrac{\,p_{k}\big(\mathbf{p}_u[n]\big)}{R_{k}\big(\mathbf{p}_u[n]\big)}
            \\
			&\text{subject to:} 
            \quad {\rm CSR}_{k}\big(\mathbf{p}_u[n],{w}_{k}[n]\big)\geq {\rm CSR}_{th}, \quad\quad\quad\quad \quad \quad \quad \quad\quad \quad\quad \quad\quad \quad \quad\ \forall n,\, \forall k,
            \label{eq:cons_CSR}
            \\
            & \quad\quad\quad\quad\quad\  F_{k}\big(\mathbf{p}_u[n],{w}_{k}[n]\big) \leq F_{th}, \quad \quad \quad\quad \quad\quad \quad \quad \quad \quad \quad \quad \quad \quad\quad \quad\quad\quad \forall n,\, \forall k,
            \label{eq:cons_outage}
			\\
			& \quad\quad\quad\quad\quad\  c_{k}\big(\mathbf{p}_u[n],{w}_{k}[n]\big)[n] \le c_{b}\big(\mathbf{p}_u[n]\big)[n], \quad\quad \quad\quad \quad \quad \quad \quad\quad \quad\quad \quad\quad\  \forall n,\, \forall k, 
            \label{eq:cons_capacity}
            \\
            & \quad\quad\quad\quad\quad\  
			0 \le p_{k}\big(\mathbf{p}_u[n]\big)[n] \le P_u^{\max},
			\,\,\, 0 \le p_{b}\big(\mathbf{p}_u[n]\big) [n]\le P_b^{\max}, \quad\quad \quad \quad \quad\  \forall n,\, \forall k,
            \label{eq:cons_pow}
			\\
			& \quad\quad\quad\quad\quad\  
			w_{min} \leq w_k[n] \leq w_{max},\,\, 0 \leq \theta_k[n] \leq \theta_{\rm FoV}/2,\,\,  0 \leq \phi_k[n]  \leq 2\,\pi, \quad\quad \forall n,\, \forall k,
            \label{eq:BW}
			\\
			& \quad\quad\quad\quad\quad\  \sqrt{((x_u[n]-x_c[n])^2+(y_u[n]-y_c[n])^2)} \leq (R_{\rm FoV}-R_{max}),  \quad\quad \quad\ \  \forall n, 
            \label{eq:cons_dim_1}
            \\
            & \quad\quad\quad\quad\quad\ \max\!\left\{
			\frac{R_{\max}+\Delta}
			{\tan\!\left(\theta_{\rm FoV}/2\right)},
			\; h^b_{min}
			\right\}
			\le h_u \le h_{\max}, \quad\quad \quad \quad \quad \quad\quad \quad\quad \quad\ \  \forall n,
            \label{eq:cons_dim_2}
		\end{align}
		\label{Pro_Form}
	\end{subequations}
	\hrulefill
	\vspace{-3mm}
\end{figure*}

\setlength{\textfloatsep}{0pt}

We present our constrained single-objective optimization problem in~\eqref{Pro_Form}. In Equation ~\eqref{Pro_Form}, the vector $\mathbf{p}_u[n]=[x_u[n],y_u[n],h_u]^{\mathsf T}$ denotes the 3D position of the UAV, relative to the world frame, at time slot $n$. The total mission duration is denoted by $T$, and is given by $
T=\sum_{n=1}^{N_t}\left(\Delta t_t[n]+\Delta t_h[n]\right)$
where $N_t$ denotes the number of UAV displacements, while $\Delta t_t[n]$ and $\Delta t_h[n]$ represent the traveling and hovering durations, respectively. The UAV propulsion energy, $E_{\mathrm{pro}}(\cdot)$, is a function of its position ${\mathbf{p}}_u[n]$ and velocity $\dot{\mathbf{p}}_u[n]$, whereas the hovering energy, $E_{\mathrm{hov}}(\cdot)$, depends only on the hovering time. The communication energy is a function of the transmit power $p_{k}(\mathbf{p}_u[n])$ and the downlink bit rate $R_{k}(\mathbf{p}_u[n])$, where $N_{b,k}$ denotes the number of bits received by the $k^{\mathrm{th}}$ user. ~\eqref{eq:cons_CSR} and \eqref{eq:cons_outage} represent the constraints associated with the quality of service (QoS) of the communication coverage, characterized by the channel secrecy rate constrained by ${\rm CSR}_{th}$, and the outage probability constrained by $F_{th}$. ~\eqref{eq:cons_capacity} and~\eqref{eq:cons_pow} represent the capacity and power constraints, respectively. The maximum transmit powers for the downlink and BS-to-UAV link are denoted by $P_k^{\max}$ and $P_b^{\max}$, respectively. ~\eqref{eq:BW} defines the beamwidth and orientation constraints, which determine the allowable dynamic ranges of the beamwidth and steering angles. Specifically, the polar orientation angle of the RF beam cannot exceed
 the half angle of the UAV field of view, i.e., $\theta_k\le \theta_{\rm FoV}/2$.
~\eqref{eq:cons_dim_1} and~\eqref{eq:cons_dim_2} represent the dimensional constraints. Specifically, the distance between the UAV position and the user cluster center, $(x_c[n],y_c[n])$, must be less than or equal to the difference between the UAV footprint radius and the maximum user group expansion radius, i.e., $R_{\rm FoV}-R_{\max}$. Furthermore, the UAV altitude, $h_u$, is constrained to satisfy a minimum operational altitude ensuring LoS with the BS, a minimum footprint requirement to cover the users, and a regulatory altitude limit imposed by the operational environment. At each time slot $n$, the UAV jointly optimizes its position and beamforming to minimize the per-slot energy consumption while satisfying the communication constraints.

\begin{algorithm}[t]
	\caption{Pseudocode of the Solution Approach}
	\label{alg:Solution_Approach}
	\begin{algorithmic}[1]
		\Require  Real-time user positions  and  feedbacks
		\Ensure UAV motion command and tuning the RF Beamforming 
		
		\State Compute FoV margins $m_{k}[n]$ and $M[n]$
		\State Evaluate QoS constraints (CSR and outage)
		\While  {A motion is detected}{}
		\If{$(M[n] \ge m_{\rm safe})$ and QoS satisfied}
		\State Hover (no motion)
		\State Check the VIT update
		\State Update beamforming parameters
		\State Check user feedbacks
		\Else
		\State Estimate motion direction $\hat{\mathbf{d}}[n]$
		\State Run the Optimizer for $d[n]$
		\State Apply the best available displacement  $d^*[n]$
		\State Update UAV position \State Update beamforming parameters
		\State Check user feedbacks
		\EndIf
		\EndWhile
	\end{algorithmic}
\end{algorithm}

\vspace{-1em}

\subsection{Solution Approach}
Due to the nonlinear and highly coupled structure of our optimization problem, obtaining a closed-form solution is challenging. Specifically, the achievable data rates depend logarithmically on the signal-to-noise ratio, which itself is a nonlinear function of the UAV position through distance-dependent pathloss and time-varying user locations. Consequently, the rate constraints introduce nonconvex feasible regions that evolve dynamically with user mobility. Moreover, the propulsion energy consumption depends on the UAV trajectory across consecutive time slots, creating temporal coupling between optimization variables. Even when the communication-energy term is neglected and the UAV altitude is fixed, the resulting trajectory optimization remains nonconvex because the logarithmic rate expressions and mobility-induced channel variations prevent analytical characterization of the optimal solution. Therefore, closed-form solutions are generally intractable, and iterative numerical optimization techniques are typically required to obtain efficient suboptimal solutions.

\label{sec:motion_detection_prediction}
In Algorithm \ref{alg:Solution_Approach}, we present our proposed solution approach, enabling a high-quality suboptimal policy. Our solution assumes a 2-D spatial domain by fixing the UAV altitude\footnote{The UAV altitude should be determined according to the maximum height of obstacles along the BS-to-UAV link, as well as the flight and aviation regulations in the deployment area. Generally, a lower altitude results in lower transmission power and reduced wind-induced turbulence; however, it also reduces the UAV footprint, and increases the risk of collisions with obstacles.}. This eliminates unnecessary maneuvers along the vertical axis, thereby reducing energy consumption. We further assume that the UAV does not predict future user trajectories. Therefore, the UAV moves toward the centroid of the user cluster as a heuristic policy. The proposed solution optimizes the displacement along this direction as follows. 

At each time step $n$, the UAV determines whether beamforming adaptation or physical repositioning is required based on the VIT signal, which indicates the proximity of users to the boundaries of the UAV’s FoV, as well as user feedback reflecting the communication quality of service (QoS). If all active users remain sufficiently within the FoV and satisfy the QoS constraints, the UAV adopts a hovering policy to minimize propulsion energy (i.e., traveling energy). Meanwhile, the UAV updates only the beamforming direction and beamwidth in response to minor changes in user positions. Otherwise, if significant changes in user positions occur, the UAV estimates the centroid of the user cluster and computes a suboptimal repositioning action that minimizes travel distance while maintaining both FoV coverage and QoS guarantees.

Let $\mathbf{p}^c_{k}[n]=[u_{k}[n],\,v_{k}[n]]^\top$ denote the pixel-coordinate vector of user $k$ in the UAV camera frame at time slot $n$. The normalized image coordinates are
\begin{equation}
\tilde{u}_{k}[n]=\frac{2\,u_{k}[n]}{u_{max}}-1,\qquad
\tilde{v}_{k}[n]=\frac{2\,v_{k}[n]}{v_{max}}-1\,,
\end{equation}
where $u_{max}$ and $v_{max}$ denote the image width and height (in pixels), respectively. The UAV's FoV boundary margin is defined as
\begin{equation}
m_{k}[n]=1-\max\left(|\tilde{u}_{k}[n]|,|\tilde{v}_{k}[n]|\right),
\end{equation}
and the worst-case visibility margin is
\begin{equation}
M[n] = \min_{k:v_{k}[n]=1} m_{k}[n]\,,
\end{equation}
where the minimum is taken over all users visible in the camera at time slot $n$, i.e., $v_{k}[n]=1$. Define the QoS feasibility indicator as
\begin{equation}
Q[n] =
\begin{cases}
1, & \text{if all users satisfy QoS constraints}, \\
0, & \text{otherwise}.
\end{cases}
\end{equation}
These quantities, $Q[n]$ and $M[n]$, act as an \emph{event-triggering mechanism} as shown in Block 9 of Fig. \ref{System_Model}. The UAV remains stationary if
\begin{equation}
(M[n] \ge M_{\rm th}) \wedge (Q[n] = 1),
\end{equation}
where $M_{\rm th}$ represents the visibility threshold and $\wedge $ denotes the logical AND operator, thereby minimizing propulsion energy. Otherwise, a motion policy is activated. The centroid of users cluster is computed as
\begin{equation}
\mathbf{g} [n]= \frac{1}{\sum{k:v_{k}[n]=1}}\sum_{{k:v_{k}[n]=1}}\mathbf{p}^w_{k}[n].
\end{equation}
The dominant motion direction is estimated as
\begin{equation}
\hat{\mathbf{d}}[n] =
\frac{\mathbf{g}[n] - \mathbf{p}_{u}[n]}{\|\mathbf{g}[n] - \mathbf{p}_{u}[n]\|}\,.
\end{equation}
The UAV motion is constrained along the dominant direction $\hat{\mathbf{d}}[n]$. The new position is computed as
\begin{equation}
\mathbf{p}_{u}[n+1] = \mathbf{p}_{u}[n] + d[n] \hat{\mathbf{d}}[n]\,,
\end{equation}
where the displacement $d[n]$ is optimized using an optimizer. Given the optimized displacement, $d^*[n]$,  the optimized position for the UAV at time $n+1$, is computed as 
\begin{equation}
\mathbf{p^*}_{u}[n+1] = \mathbf{p}_{u}[n] + d^*[n] \hat{\mathbf{d}}[n]\,,
\end{equation}


\begin{algorithm}[t]
	\footnotesize
	\caption{Pseudocode of LS, BS, and GS Optimizers}
	\label{alg:Optimizers}
	\begin{algorithmic}[1]
		\State \textbf{Optimizer 1: Binary Search (BS)}
		
		\State Initialize two random indices:
		$i_{\rm min}\geq 1$ and $i_{\rm max}\leq N_{\delta}$.
		
		\State Check the feasibility of
		$d[n]=\{i_{\rm min}\times \delta,\; i_{\rm max}\times \delta\}$.
		
		\While{$j \leq J$}
		
		\If{only one index is feasible}
		
		\State Evaluate the feasibility of
		$d[n]=i_{\rm mid}\times \delta$, where
		$
		i_{\rm mid}=(i_{\rm min}+i_{\rm max})/2
		$
		
		\If{feasible}
		\State feasible index $\leftarrow i_{\rm mid}$
		\Else
		\State infeasible index $\leftarrow i_{\rm mid}$
		\EndIf
		
		\ElsIf{none is feasible}
		
		\State
		$i_{\rm min}=i_{\rm max}$ and
		$i_{\rm max}\in[i_{\rm min}+1,N_{\delta}]$
		
		\Else
		
		\State
		$i_{\rm max}=i_{\rm min}$ and
		$i_{\rm min}\in[1,i_{\rm max}-1]$
		
		\EndIf
		
		\State $j=j+1$
		
		\EndWhile
		
		\State Check the feasibility of
		$d[n]=\{i_{\rm min}\times \delta,\; i_{\rm max}\times \delta\}$
		
		\If{a feasible solution is found}
		\State
		$d^*[n]=$ minimum feasible value of
		$\{i_{\rm min}\times \delta,\; i_{\rm max}\times \delta\}$
		\Else
		\State $d^*[n]=N_{\delta}\times \delta$
		\EndIf

		\Statex
		\State \textbf{Optimizer 2: Linear Search (LS)}
		
		\For{$j=1:N_{\delta}$}
		\State Check the feasibility of $d[n]=j \times \delta$
		\If{feasible}
		\State $d^*[n]=d[n]$
		\EndIf
		\EndFor
				
				\Statex
		
		\State \textbf{Optimizer 3: Genetic Search (GS)}
		
		\State Initialize the GS population:
	$
		\delta \leq d[n] \leq
		\|\mathbf{g}[n]-\mathbf{q}_{u}[n]\|
	$
		
		\For{$G = 1:G_{\max}$}
		
		\State Check the feasibility of the population:
		$\{d_1[n],\dots,d_{N_{\rm pop}}[n]\}$
				\State Select minimum-displacement parent solutions
				
				\State Apply crossover and mutation operations
				
				\State Preserve elite solutions

		\EndFor
		
		\If{ feasibility is found}
		
		\State
		$d^*[n]=
		\min\{d_1[n],\dots,d_{N_{\rm pop}}[n]\}$
		
		\Else
		
		\State
		$d^*[n]=
		\|\mathbf{g}[n]-\mathbf{q}_{u}[n]\|$
		
		\EndIf
		
	\end{algorithmic}
\end{algorithm}

\vspace{-1.2em}
\subsection{Searching-Based Optimizers}
\label{Optimizer}
To satisfy the low-complexity, real-time requirements of the considered application and exploiting the monotonic feasibility property along the displacement direction, we employ a binary search (BS) method. For comparison, linear search (LS) and genetic search (GS) methods are also considered \cite{Goldberg_GA}. All optimizers solve ~\eqref{Pro_Form} according to Algorithm~\ref{alg:Solution_Approach}, while Algorithm~\ref{alg:Optimizers} summarizes their implementations for determining the near-optimal displacement $d^{*}[n]$.

BS iteratively reduces the search interval,
$0 < d[n]\leq \|\mathbf{g}[n]-\mathbf{p}_{u}[n]\|$,
by evaluating the midpoint candidate and updating the lower and upper bounds according to feasibility until convergence. LS discretizes the search interval using a step size $\delta$ and sequentially evaluates candidate displacements. Owing to the monotonic feasibility property, the first feasible candidate is selected as the solution. GS initializes a population of candidate displacements within the feasible interval and iteratively refines the population through selection, crossover, mutation, and elitism until convergence.

The computational time of all three optimizers can be expressed as
$
\mathcal{O}\left(J_x\mathcal{O}_{f}\right), \quad x\in\{B,L,G\},
$
where $\mathcal{O}_{f}$ denotes the complexity of a single cost-function evaluation, and $J_B$, $J_L$, and $J_G$ represent the numbers of iterations required by BS, LS, and GS, respectively. For LS, $J_L=\lfloor \|\mathbf{g}[n]-\mathbf{p}_{u}[n]\|/\delta \rfloor$, while for GS, $J_G$ equals the population size multiplied by the number of generations. Owing to the logarithmic reduction of the search interval, BS typically requires significantly fewer iterations to achieve near-optimal performance, i.e.,
$
J_B \ll \min\{J_L,J_G\}.
$

\setlength{\textfloatsep}{30pt}


\begin{table}[t]
	\centering
	\caption{Simulation Parameters \cite{2014_PathLoss, 2019_Pwer_Hovering_Model, Simulation_Parameters_1, A_New_Framework}.}
	\begin{tabular}
    {p{0.46\columnwidth}p{0.16\columnwidth}p{0.23\columnwidth}}
		\hline
		Parameter & Symbol  & Value \\ 
		\hline
		CSR Threshold & ${\rm CRS}_{th}$ & 5 bits/channel-use \\
		Capacity Threshold & $c_{th}$ & 5 bits/channel-use \\
		Outage Threshold & $F_{th}$ & 0.05 \\
		Displacement Variances in X and Y & $\sigma_{x}^2$ and $\sigma_{y}^2$ & 0.25 $m^2$ \\
		Safety-zone Radius & $R_{\mathrm{s}}$ & 5 m \\
		UAV Altitude & $h_u$ & 20 m \\
		Number of Users & $K$ & 5 \\
		User Height & $h_k$ & 1.5 m \\
		Minimum/maximum Group Radius & $R_{\min}, R_{\max}$ & 20 m \\
		UAV Maximum Transmit Power \footnote{In Canada, the  Innovation, Science and Economic Development Canada (ISED) limits the maximum power to 4 Watt.} & $P_u^{\max}$ & 3 W \\
		FOV Safety Margin & $\Delta$ & $20$ m \\
		UAV Footprint Radius & $R_{\rm FoV}$ & $40$ m\\
		Localization Error in X and Y & $\sigma_x$ and $\sigma_y$ & 0.25 m \\
		Receiver Aperture Radius & $A_r$ & 0.1 m \\
		Carrier Frequency & $f_c$ & $2.4\times10^{9}$ Hz \\
		Speed of Light & $c$ & $3\times10^{8}$ m/s \\
		downlink Bandwidth & $B_w$ & $20\times10^{6}$ Hz \\
		UAV Travel Speed & $V_{\text{travel}}$ & 10 m/s \\
		\hline
	\end{tabular}
	\label{Sim_Par}
    \vspace{-3.3em}
\end{table}

\begin{figure*}[t]
	\centering
	
	\begin{subfigure}[b]{0.45\textwidth}
		\centering
		\includegraphics[width=\linewidth,height=6cm]{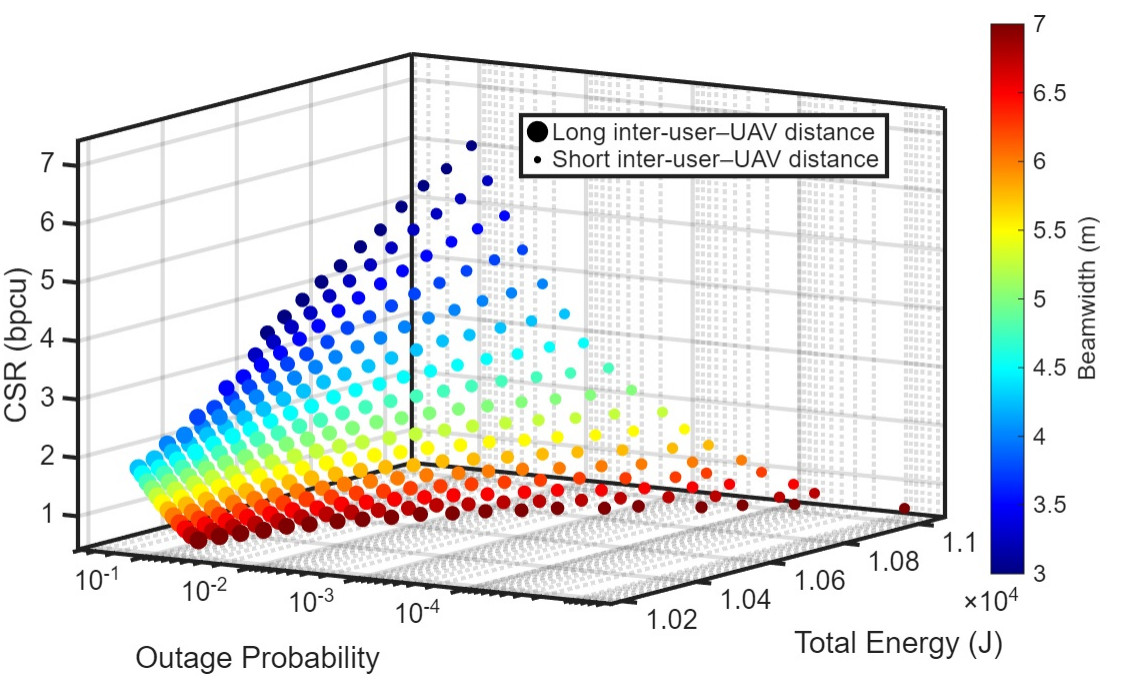}
		\caption{}
		\label{Outage_CSR_Energy}
	\end{subfigure}
	\hfill
		\begin{subfigure}[b]{0.45\textwidth}
			\centering
			\includegraphics[width=\linewidth,height=6cm]{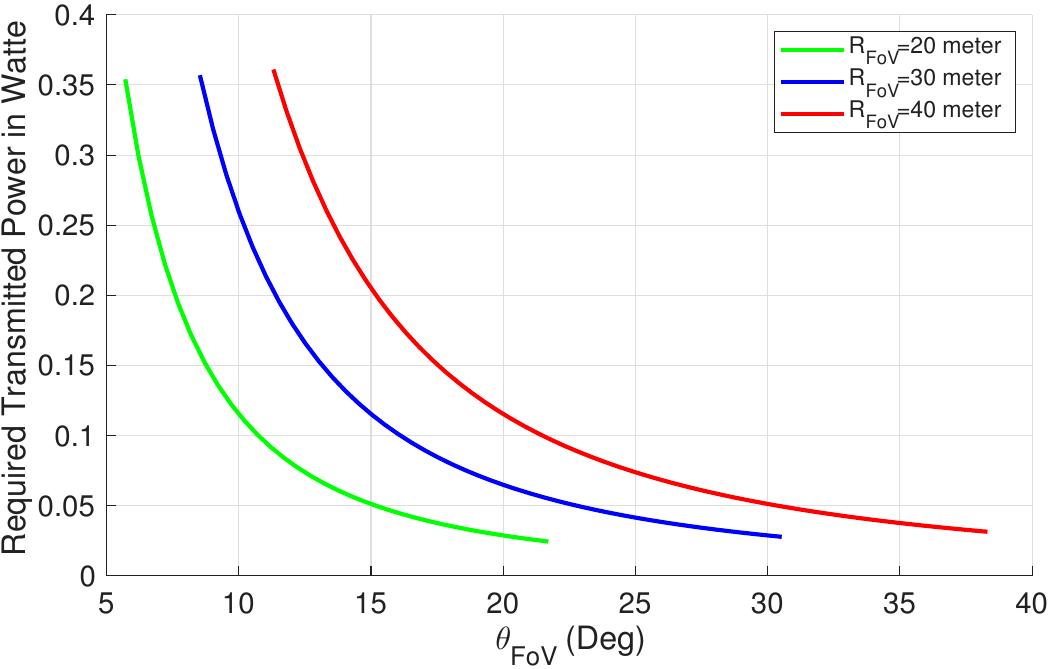}
			\caption{}
			\label{Power_FoV}
		\end{subfigure}
        \vspace{-0.8em}
	\caption{
		Performance analysis: (a) Tradeoff among energy consumption, CSR, and outage probability for different $h_u$ and $w$. (b) Required transmit power versus FoV.
	}
	\label{fig:power_tradeoff}
    \vspace{-1em}
\end{figure*}

\begin{figure*}[t]
	\centering
	
	\begin{subfigure}[b]{0.47\textwidth}
		\centering
		\includegraphics[width=.99\linewidth]{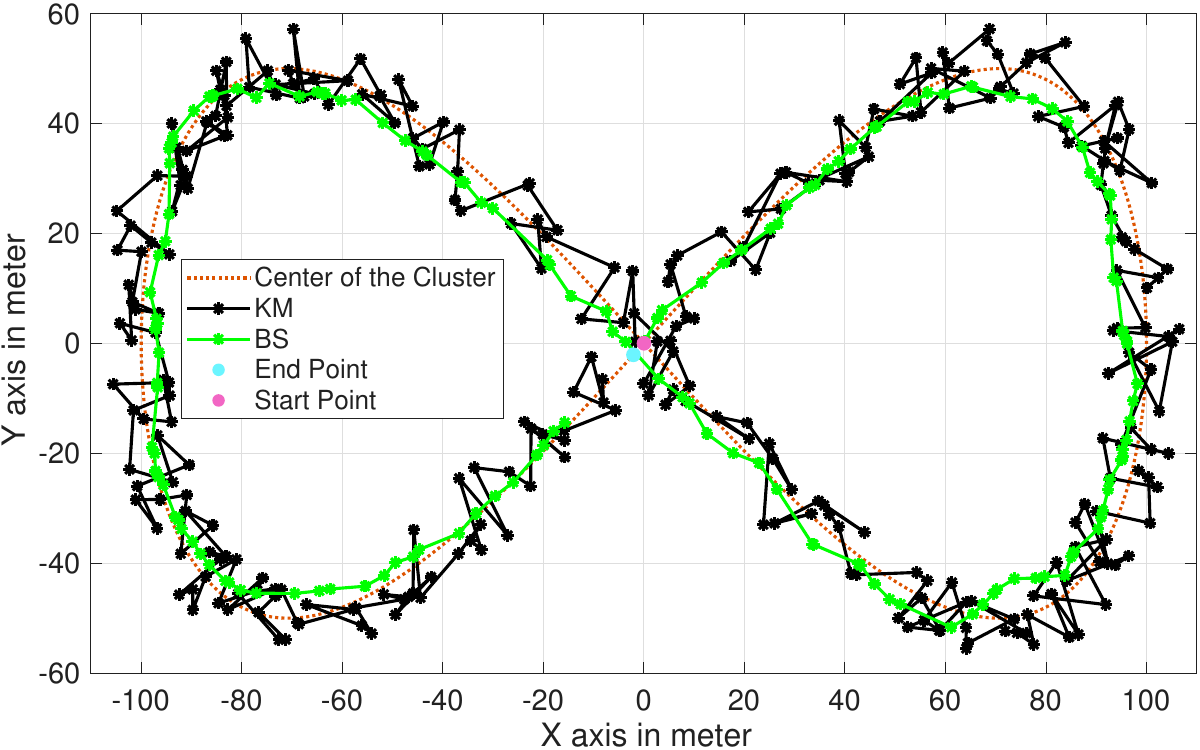}
		\caption{}
		\label{Path_BS_KM}
	\end{subfigure}
	\hfill
	\begin{subfigure}[b]{0.47\textwidth}
		\centering
		\includegraphics[width=.99\linewidth]{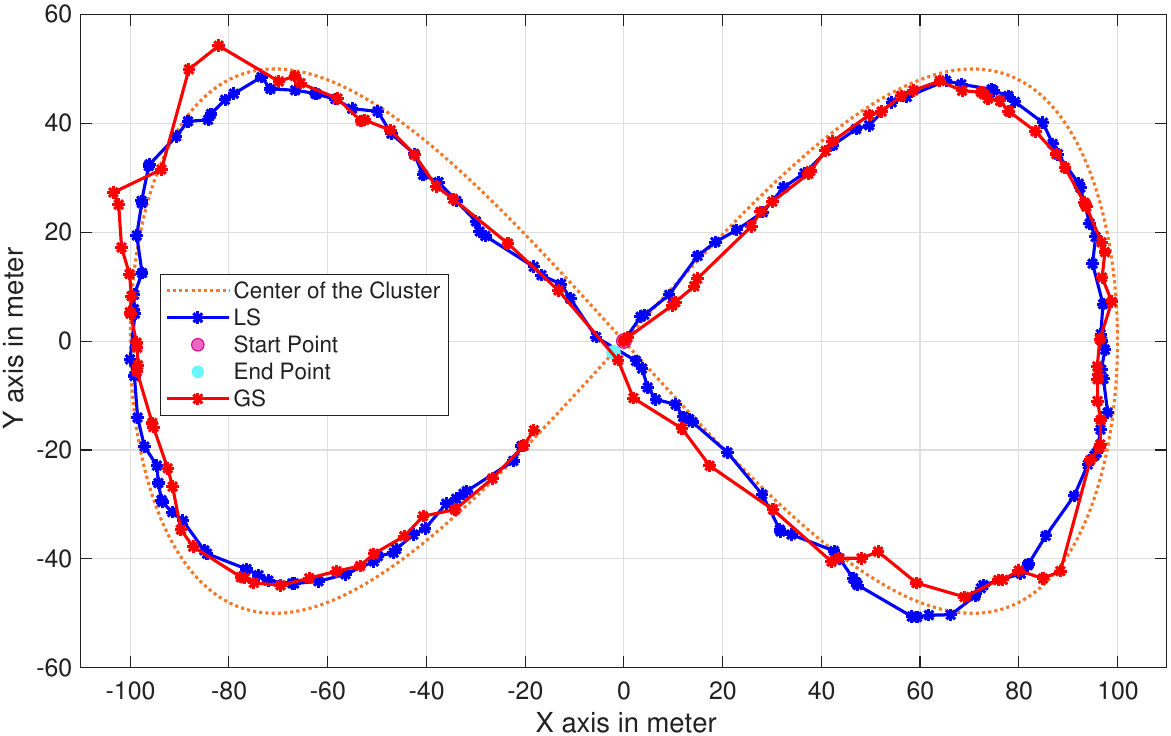}
		\caption{}
		\label{Path_LS_GA}
	\end{subfigure}
	\vspace{-0.8em}
	\caption{UAV trajectory paths for the BS and KM optimizers in (a), and for the LS and GS optimizers in (b).}
	\label{Trajectory_Path}
    \vspace{-1.5em}
\end{figure*}

\section{Numerical Results}
\label{Numerical_Results}
This section evaluates the proposed VIT-UAVCom system in terms of total energy consumption, energy efficiency, outage probability, and CSR. The optimization problem in ~\eqref{Pro_Form} is solved using the proposed framework in Algorithm~\ref{alg:Solution_Approach} with the BS, LS, and GS optimizers described in Algorithm~\ref{alg:Optimizers}. Performance is compared against a K-means (KM) benchmark, where the UAV follows the centroid of the user cluster without applying the proposed optimization framework. Simulations consider a dynamic mission in which a UAV serves a group of mobile users over varying mission durations. The simulation parameters are summarized in Table~\ref{Sim_Par} \cite{2014_PathLoss, 2019_Pwer_Hovering_Model, Simulation_Parameters_1, A_New_Framework}.

\begin{figure*}[t]
	\centering
	
	\begin{subfigure}[b]{0.45\textwidth}
		\centering
		\includegraphics[width=\textwidth]{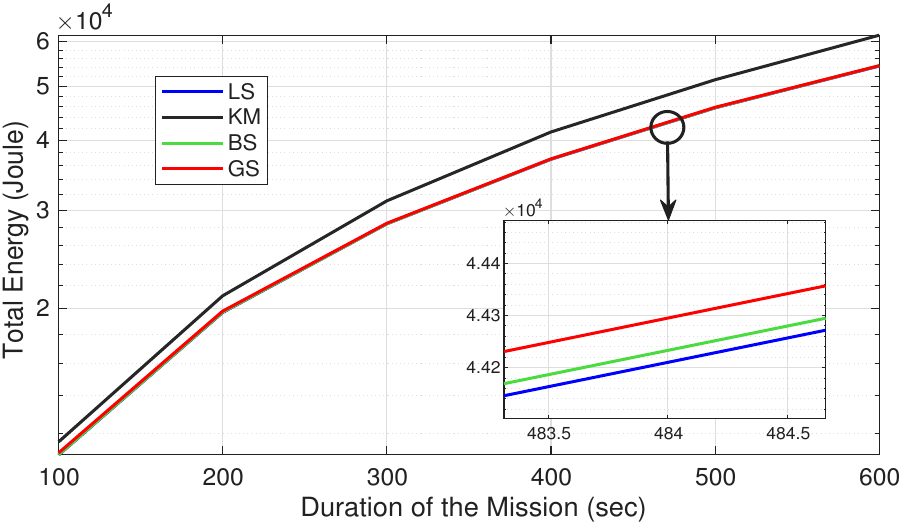}
		\caption{}
		\label{fig:sub1}
        \vspace{-1em}
	\end{subfigure}
	\hfill
	\begin{subfigure}[b]{0.45\textwidth}
		\centering
		\includegraphics[width=\textwidth]{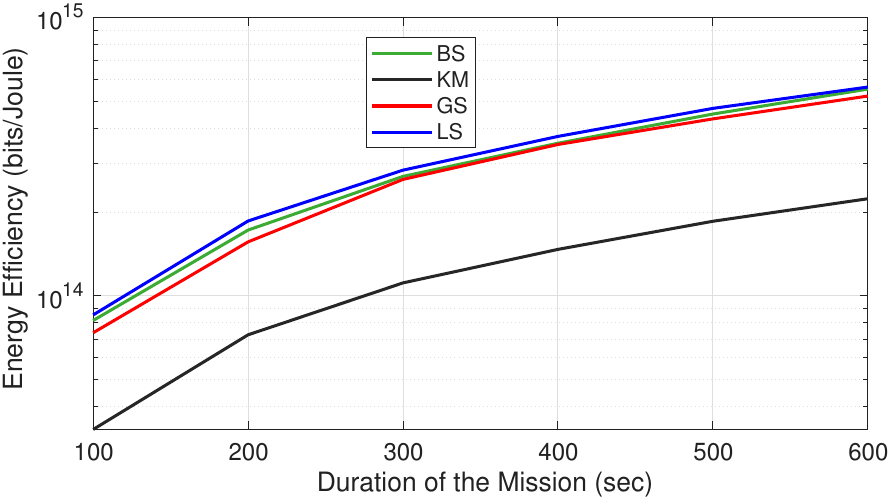}
		\caption{}
		\label{fig:sub2}
        \vspace{-1em}
	\end{subfigure}
	
	\vspace{0.4cm}
	
	\begin{subfigure}[b]{0.45\textwidth}
		\centering
		\includegraphics[width=\textwidth]{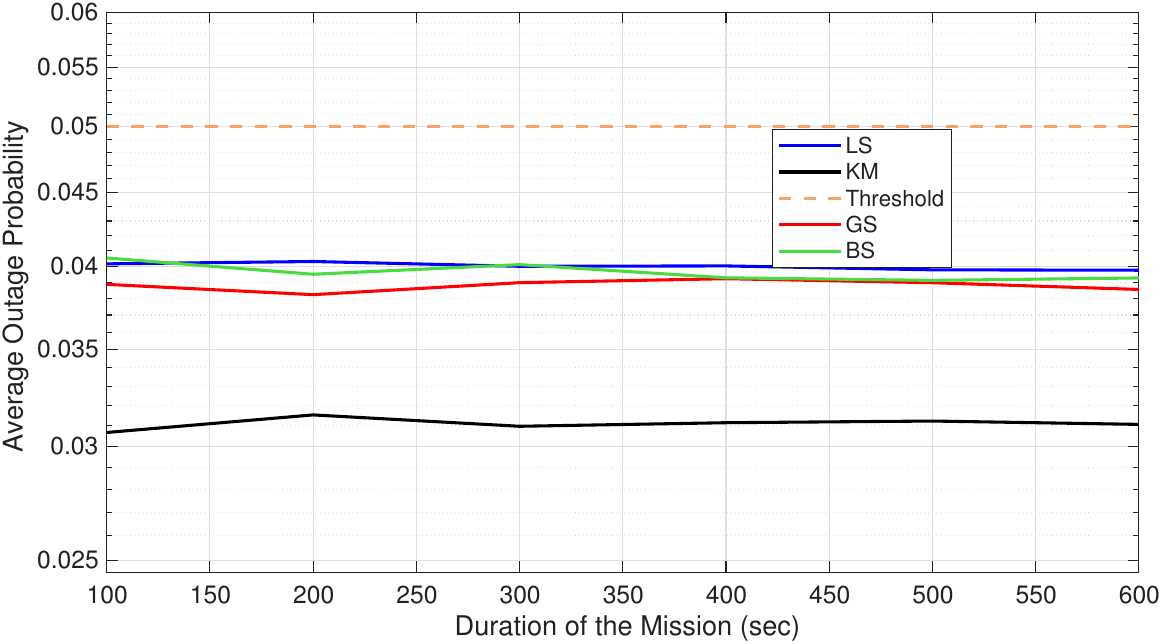}
		\caption{}
		\label{fig:sub3}
	\end{subfigure}
	\hfill
	\begin{subfigure}[b]{0.45\textwidth}
		\centering
		\includegraphics[width=\textwidth]{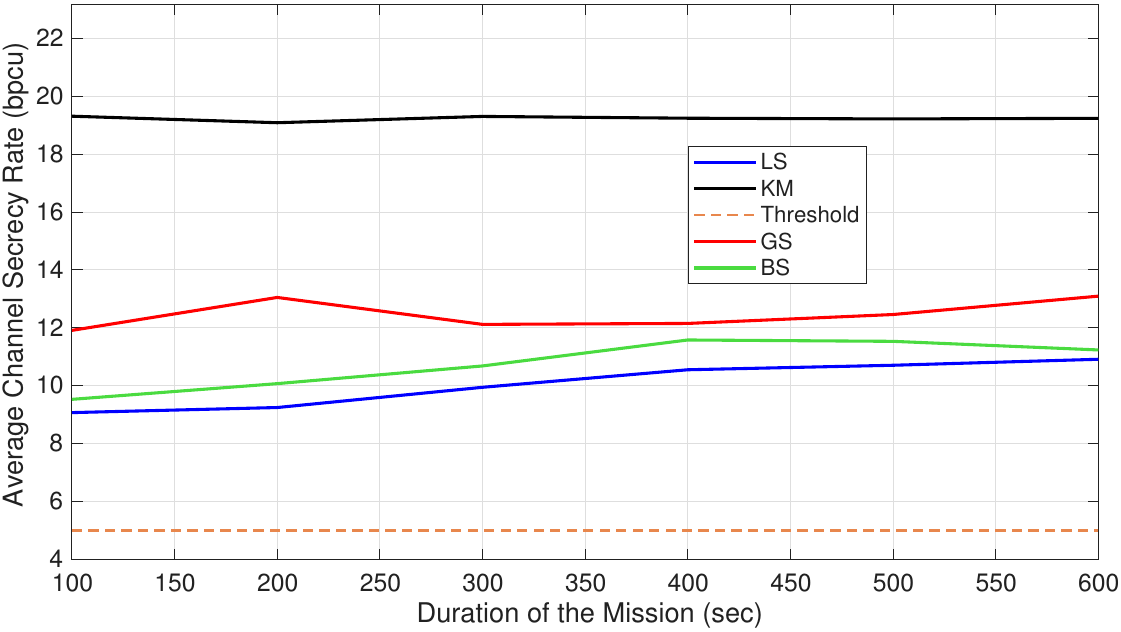}
		\caption{}
		\label{fig:sub4}
	\end{subfigure}
	
	\caption{
		Performance versus mission duration: (a) total energy consumption, (b) EE, (c) outage probability, and (d) CSR.
	}
	\label{fig:mission_results}
    \vspace{-1em}
\end{figure*}

\vspace{-0.5em}
\subsection{System Characteristics}

Fig.~\ref{fig:power_tradeoff} illustrates the fundamental tradeoffs among propulsion  energy, CSR, outage probability, transmit power, UAV altitude, and FoV.

In Fig.~\ref{fig:power_tradeoff}(\subref{Outage_CSR_Energy}), the UAV initially hovers at $(0,0,300)$ and may move closer to a user located at $(0,0,0)$. The figure shows the relationship among propulsion  energy, CSR, and outage probability for different UAV altitudes, $h_u=\{100,110,120,140,\ldots,300\}$~m, and beamwidths, $w_k=\{3,3.25,3.5,3.75,\ldots,7\}$~m. Each point represents a design pair ${h_u,w_k}$, where color indicates beamwidth and marker size indicates propulsion  distance. As observed, narrow beams generally improve CSR by enhancing the legitimate link while reducing signal leakage toward the eavesdropper. However, they are more sensitive to wind- and mobility-induced misalignment, resulting in higher outage probability. In contrast, wider beams improve robustness and reduce outage probability at the expense of lower CSR. Similarly, reducing UAV altitude decreases path loss and beam misalignment, thereby improving both CSR and outage performance, but requires additional propulsion  energy due to UAV movement. These observations reveal that maximizing CSR, minimizing outage probability, and minimizing propulsion  energy are conflicting objectives, requiring a balanced operating point.

Fig.~\ref{fig:power_tradeoff}(\subref{Power_FoV}) shows the required transmit power versus the FoV angle, $\theta_{\rm FoV}$, for different footprint radii, $R_{\rm FoV}$. For a given footprint, the required transmit power decreases as $\theta_{\rm FoV}$ increases because a larger FoV allows the footprint to be achieved at a lower altitude according to $R_{\rm FoV}=h_u\tan(\theta_{\rm FoV}/2)$, thereby reducing path loss. Conversely, larger footprint radii require higher transmit power, shifting the curves upward. Moreover, the power reduction becomes marginal at moderate-to-large FoV angles, indicating an operating region that balances footprint coverage and power efficiency. Overall, the results highlight the interdependent tradeoffs among UAV altitude, footprint radius, beamwidth, and transmit power.

\vspace{-1.2em}
\subsection{System Performance}
\vspace{-0.3em}
Fig.~\ref{Trajectory_Path} illustrates the UAV trajectories generated by the considered optimization methods and the KM benchmark. The KM benchmark produces irregular trajectories with frequent direction changes due to continuously following the cluster center, resulting in higher travel distance and propulsion energy consumption. In contrast, the proposed optimizers generate smoother and more structured paths that better adapt to user mobility. Among them, LS achieves the smoothest trajectory, followed by BS and GS, contributing to improved energy efficiency and communication performance.

Figures~\ref{fig:mission_results}(\subref{fig:sub1})--(\subref{fig:sub4}) compare the optimizers in terms of total energy consumption, energy efficiency (EE), outage probability, and CSR versus mission duration. As shown in Fig.~\ref{fig:mission_results}(\subref{fig:sub1}), energy consumption increases with mission duration for all methods. However, the proposed optimizers outperform KM, with LS achieving the lowest energy consumption. At $T=500$ s, the optimizers consume approximately $4.5\times10^4$ J compared with $5\times10^4$ J for KM, corresponding to an energy reduction of about $10\%$.

Fig.~\ref{fig:mission_results}(\subref{fig:sub2}) shows that EE increases with mission duration as the UAV spends more time hovering than moving. At $T=500$ s, the proposed optimizers achieve approximately $4.4\times10^{14}$ bpJ, compared with $1.8\times10^{14}$ bpJ for KM, yielding an EE improvement of about $144\%$.

Figs.~\ref{fig:mission_results}(\subref{fig:sub3}) and (\subref{fig:sub4}) show that outage probability and CSR remain relatively stable across mission durations. All proposed methods satisfy the outage and CSR requirements. KM achieves the lowest outage probability and highest CSR because it remains closer to the users. At $T=500$ s, LS, BS, and GS achieve CSR values of approximately $13$, $11$, and $9$ bpcu, respectively, compared with $19$ bpcu for KM. 

In terms of computational time, BS requires only $10$ iterations and approximately $12$ minutes of execution time, compared with $50$ iterations and $25$ minutes for LS, and $50$ iterations and $40$ minutes for GS. Therefore, BS reduces execution time by about $50\%$ relative to LS, while GS incurs the highest computational time and the lowest overall performance.

\vspace{-1em}
\section{Conclusions and Discussions}
\label{Conclusion}
This paper presented a VIT-UAVCom framework that improves UAV energy efficiency while ensuring reliable communication in GPS-denied environments using onboard camera and IMU sensors. By integrating vision–inertial tracking with communication design, the system removes GPS dependency and enhances privacy and security. Among the optimizers, BS achieved the lowest computational time, while LS yielded the highest energy efficiency. Overall, the framework supports lightweight real-time optimization for green NTN communications with reduced computational complexity. Future work includes mobility prediction, machine learning-based optimization, multi-sensor fusion, and robustness improvements in challenging scenarios such as low visibility, nighttime, and cluttered environments.

\vspace{-1em}

\bibliographystyle{IEEEtran}
\bibliography{Abbreviated_References_V2}

\end{document}